\documentclass[12pt]{article}

\def\journal#1#2#3#4{{\it #1} {\bf #2} (#3) #4}
\def\epj{Euro. Phys. Jour.}
\def\prl{Phys. Rev. Lett.}
\def\pl{Phys. Lett.}
\def\np{Nucl. Phys.}
\def\ptp{Prog. Theor. Phys.}

\def\zp{Z. Phys.}
\def\pr{Phys. Rev.}

\def\jhep{JHEP}

\def\ppnp{Prog. Part. Nucl. Phys.}
\def\ml{{\hat{m_{\ell}}}}
\def\ddp{{D^\prime}}
\def\mc{{\hat{m_c}}}
\usepackage{epsfig}

\def\o{{\cal O}}
\def\c{C}
\def\cs{{\c_7}}
\def\cn{{\c_9}}
\def\ct{{\c_{10}}}
\def\cne{\cn^{\rm eff}}
\def\cse{\cs^{\rm eff}}
\def\m{{\cal M}}
\def\gl{\Gamma}
\def\g{\gamma}
\def\l{\ell}

\def\lb{\bar{\l}}

\def\ph{{p_h}}

\def\he{{\cal H}_{\rm eff}}

\def\d{{\rm d}}

\def\t{{\rm T}}

\def\mh{\hat{m}}
\def\mbh{\mh_b}
\def\mph{\mh_K}
\def\mvh{\mh_{K^*}}

\def\mlh{\mh_\l}
\def\qh{\hat{q}}
\def\pvh{\hat{p}_{K^*}}

\def\pbh{\hat{p}_B}
\def\ph{\hat{p}}
\def\sh{\hat{s}}

\def\t{{\cal T}}
\def\a{{\cal A}}
\def\ep{{\epsilon^\ast}}
\def\ap{{A^\prime}}
\def\bp{{B^\prime}}
\def\cp{{C^\prime}}
\def\rp{{D^\prime}}
\def\uh{{\hat{u}}}
\def\la{{\lambda}}

\def\be{\begin{equation}}
\def\ee{\end{equation}}
\def\ba{\begin{eqnarray}}
\def\ea{\end{eqnarray}}
\begin{document}
\renewcommand{\thefootnote}{\fnsymbol{footnote}}

\begin{titlepage}

\begin{flushright}
\begin{tabular}{l}
DESY 99--146\\
CERN--TH/99--298\\
LNF--99/026 (P)\\
SLAC-PUB-8269\\
October 1999
\end{tabular}
\end{flushright}
\vskip0.5cm
\begin{center}
{\LARGE\bf
A Comparative Study of the Decays $B \to (K,K^*) \ell^+ \ell^-$ in Standard 
Model and Supersymmetric Theories}\footnote{Work supported by the
Department of Energy, Contract DE-AC03-76SF00515}

\vspace*{0.5cm}

       {\bf A. Ali}$^{1,}$\footnote{
        E-mail : ali@x4u2.desy.de},
        {\bf Patricia Ball}$^{2,}$\footnote{
        E-mail : Patricia.Ball@cern.ch}, 
        {\bf L. T. Handoko}$^{1,3,}$\footnote{
        E-mail : handoko@mail.desy.de}
        and  
        {\bf G. Hiller}$^{4,}$\footnote{
        E-mail : Gudrun.Hiller@lnf.infn.it; Address since Oct.1, 1999:
        SLAC, P.O.Box 4349, Stanford, CA 94309.}

        \vspace{0.5cm}

        $^1$Deutsches Elektronen-Synchrotron DESY, \\
        Notkestr. 85, D--22607 Hamburg, Germany 

        \bigskip

        $^2$ CERN/TH, CH--1211 Geneva 23, Switzerland

\bigskip

        $^3$Lab. for Theoretical Physics and Mathematics, LIPI, \\
        Kom. Puspitek Serpong P3FT--LIPI, Tangerang 15310, Indonesia

        \bigskip

        $^4$INFN, Laboratori Nazionali di Frascati, \\
        P.O. Box 13, I--00044 Frascati, Italy




\bigskip

  {\bf Abstract\\[10pt]}
\end{center}
\noindent
Using improved theoretical calculations of the decay form factors in
the Light Cone-QCD sum rule approach,
we investigate the decay rates, dilepton invariant mass spectra and 
the forward-backward (FB) asymmetry in the decays  $B \to (K,K^*) \ell^+ 
\ell^-$ ($\ell^\pm =e^\pm,\mu^\pm,\tau^\pm$) in the standard  model (SM) 
and a number of popular variants of 
the supersymmetric (SUSY) models. Theoretical precision on the 
differential decay rates and FB-asymmetry is estimated in these 
theories taking into account various parametric uncertainties.
We show that existing data on $B \to
X_s \gamma$ and the experimental upper limit on the branching ratio
${\cal B}(B \to K^* \mu^+ \mu^-)$ provide interesting bounds on the 
coefficients of the underlying effective theory. We argue
that the FB-asymmetry in $B \to K^* \ell^+ \ell^-$ constitutes a 
precision test of the SM and its measurement in forthcoming experiments may 
reveal new physics. In 
particular, the presently allowed large-$\tan \beta$ solutions in SUGRA 
models, as well as more general flavor-violating SUSY models, yield 
FB-asymmetries  which are characteristically different from the 
corresponding ones in the SM.

\vfill

\centerline{\em submitted to Physical Review D}

\bigskip

\centerline{{\sc Pacs} numbers: 13.20.He, 13.25.Hw}
\end{titlepage}

\renewcommand{\thefootnote}{\arabic{footnote}}
\setcounter{footnote}{0}

\section{\bf Introduction}
  The flavor-changing-neutral-current (FCNC) transitions $B \to (X_s,X_d) 
\gamma$ and $B \to (X_s,X_d) \ell^+ \ell^-$, with $X_s(X_d)$ being
hadrons with overall strangeness $S=\pm 1(S=0)$, provide potentially 
stringent 
tests of the SM in flavor physics. FCNC transitions are forbidden 
in the SM Lagrangian and are induced by the 
GIM-amplitudes \cite{GIM} at the loop level, which makes their
effective strengths small. In addition, these transitions may also be 
parametrically
suppressed in the SM due to their dependence on the weak mixing angles
of the quark-flavor
rotation matrix --- the Cabibbo-Kobayashi-Maskawa matrix $V_{\mbox{CKM}}$
\cite{CKM}. These two circumstances make the FCNC decays relatively rare  
and hence vulnerable to the presence of new physics.
In the context of the SM, the potential interest in rare $B$-decays  
is that they would provide a quantitative determination of the quark-flavor 
rotation matrix, in particular the matrix elements $V_{tb}$, $V_{td}$ 
and $V_{ts}$ \cite{agm93,kms97,ah98-1,aag98}. A beginning in that 
direction has already been made by 
the measurement of the branching ratio ${\cal B}(B \to X_s \gamma)$ 
\cite{cleobsg,alephbsg}, yielding $\vert V_{ts}V_{tb}^*\vert =0.035 \pm 
0.004$ \cite{alexander98}, in agreement with the expectations based on the 
CKM-unitarity \cite{alisb97}. Since complementary information will also be 
available
from the $B_s^0$-$\overline{B_s^0}$- and $B_d^0$-$\overline{B_d^0}$-mixing 
induced mass differences $\Delta M_s$ and $\Delta M_d$, respectively, 
and from a number of rare kaon decays \cite{Buraskaon}, the parameters of 
the CKM matrix, which are already fairly constrained in the SM 
\cite{al99,mele99,prs99}, will
be multiply determined. This will result either in precise determination 
of the SM parameters in the flavor sector, comparable to the 
precision of the electroweak parameters of the SM \cite{pdg98}, or, 
more optimistically, in the discovery of new physics. Thus, FCNC 
processes are potentially effective tools in searching for 
new physics, with the supersymmetric theories receiving special attention
in this context 
\cite{al99,bertolini,agm95,cmw,hw97,goto96,MFVbsg,goto99,LMSS99,bksusy}.
 
Inclusive decay rates and distributions are relatively robust theoretically,
making them well-suited to search for new physics which may result
in distortions of the SM-distributions. Concerning rare $B$-decays, we
recall that the 
shape of the photon energy spectra in the radiative decays $B \to (X_s,X_d) 
\gamma$ depends on the underlying physics. However, deviations from the
SM-based normalized photon-energy distributions are expected only for the
low-to-intermediate photon energies, where the individual contributions
from the various operators in the underlying effective theory are 
comparable. 
Measuring the low-$E_\gamma$ spectrum is, however, a formidable task in the 
present experimental set-up. More promising from the point of view of 
observing new-physics-induced distortions in the distributions are the
decays $B \to (X_s,X_d) \ell^+ \ell^-$, which  provide the possibility of
measuring Dalitz-distributions in a number of variables, which in turn could
be used to determine the coefficients of the effective vertices in the 
underlying theory \cite{agm95}. This program is
somewhat handicapped by the fact that heavy quark expansion
in $1/m_b$ breaks down near the end-points of the spectra 
\cite{ahhm96,bi98}, near the $c\bar{c}$-threshold and in the resonant 
region. Thus, a certain amount of modeling is unavoidable for the complete
phenomenological profile of the decays $B \to X_s \ell^+ \ell^-$. A number
of studies has been undertaken to assess the 
non-perturbative effects 
\cite{ahhm96,bir98,amm91,ks96,melikhov,lsw98,ah98-2}, allowing to define 
limited kinematic regions where the short-distance physics in the SM and 
alternative theories can be quantitatively studied.   

 While the inclusive rare decays discussed above are theoretically cleaner 
than exclusive decays, which require additionally the knowledge of form 
factors, they are also more difficult to measure. Present best 
limits from the CLEO collaboration on $B \to X_s \mu^+ \mu^-$ and
$B \to X_s e^+ e^-$ \cite{cleobsll} decays are typically an order of 
magnitude larger than the corresponding SM-based estimates 
\cite{ahhm96}. Moreover, inclusive rare decays are a
challange for experiments operating at hadron machines.
However, it is encouraging that the FCNC exclusive 
semileptonic decays, in particular the $B 
\to (K,K^*) \mu^+ \mu^-$ modes, are accessible to a wider variety of
experiments. As we will argue quantitatively in this paper, some of
the present experimental bounds on these (and related $e^+ e^-$ modes)
\cite{cdfexcl,cleoexcl} are already  quite stringent. With the advent of 
the Fermilab booster, HERA-B, 
experiments at the LHC, and also the ongoing experiments at 
CLEO and the B-factories, the decays of interest $B \to (K,K^*) \ell^+ \ell^-$
will be precisely measured. It is therefore worthwhile to return to a 
comparative study of these decays in the SM and some candidate theories
of physics beyond the SM to ascertain if these modes could be meaningfully
used for searches of beyond-the-SM physics.

 In the context of the SM, exclusive FCNC semileptonic $B$-decays have been
studied in a number of papers  
\cite{deshpande,dpr90,am92,giw94,mn96,roberts,geng-kao,colangelo,burdman} 
with varying degrees of 
theoretical rigor and emphasis. The main purpose of this paper is twofold:
First, we would like to report on an improved calculation of the decay
form factors using the technique of the Light cone-QCD sum rules (LCSR)   
\cite{BBK,chernB}. Early studies of exclusive $B$ decays in the LCSR
approach were restricted to contributions of leading twist and
did not take into account radiative corrections (see
Refs.~\cite{VMBreview,KRreview} for a review and references
to original publications). In the present paper, we use the results of
\cite{ballbraun98} for vector form factors, which include NLO radiative
corrections and higher twist corrections up to twist 4
\cite{BBKT,NPB99}. For $B\to K$ form factors we improve on the results
obtained in \cite{ball98} by including the twist 4 mass correction
terms calculated in \cite{JHEP99}. Second, we apply this technology to
the SM and some popular variants of the SUSY models to determine the 
phenomenological
profiles of the decays $B \to (K,K^*) \ell^+ \ell^-$ in these
scenarios. For the latter, we choose minimal- and non-minimal SUGRA 
models, minimal flavor violating  supersymmetric (MFV) model, and a 
general flavor-violating  supersymmetric framework, studied in the mass 
insertion approximation (MIA). 
While all these models have been studied quite extensively
in the literature for the inclusive decays $B \to X_s \gamma$ and
$B \to X_s \ell^+ \ell^-$ 
\cite{bertolini,agm95,cmw,hw97,goto96,MFVbsg,goto99,LMSS99,bksusy},
 we are not aware of corresponding studies for the exclusive
decays. We strive to fill this gap in this paper.
 
 With our goals clearly stated, we turn to the main issues in
the inclusive and exclusive rare $B$-decays. 
 Using the language of effective theories
and restricting ourselves to the SM and SUSY, 
the short-distance contributions in the decays $B \to X_s \gamma$ and $B \to 
X_s \ell^+ \ell^-$, and the exclusive decays of interest to us, are 
determined by three coefficients, called  $\cse$, $\cn$ and $\ct$  
\cite{burasmuenz,misiakE}.\footnote{In general, more operators are
present in supersymmetric theories and we discuss their possible effects
later in this paper.} Of 
these, $\vert \cse 
\vert$ --- the modulus of the effective coefficient of the electromagnetic 
penguin operator 
--- is bounded by the present experimental measurements of the $B \to 
X_s \gamma$ branching ratio \cite{cleobsg,alephbsg}. Using the 95\% C.L.
upper and lower bounds from the updated CLEO measurements \cite{cleobsg}:
\be  
2.0 \times 10^{-4} < {\cal B}(B \to X_s \gamma) < 4.5 \times 10^{-4} ~,
\label{bsgamcleo}
\ee
one gets in the next-to-leading precision the bounds, 
\be
0.28 \leq \vert \cse(m_B)\vert \leq 0.41~.
\label{c7effbound}
\ee  
The magnitude of $\cse(m_B)$ in the SM \cite{misiaketal97} is well
within the CLEO bounds but there is no experimental information on the
phase of $\cse(m_B)$. It is imperative to determine this sign
experimentally, as it is model-dependent.
In particular, in SUGRA-type theories,
both positive and negative-valued solutions for $\cse(m_B)$ are allowed
in different SUSY-parameter regions.

 Despite the present lack of direct information on the sign of $\cse(m_B)$, 
the bound in Eq.~(\ref{c7effbound}) is quite stringent
and effectively limits possible new-physics effects due to the inherent
correlations among the branching ratio ${\cal B}(B \to X_s \gamma)$
and other observable quantities such as the $B^0$-$\overline{B^0}$ mixing,
$\epsilon_K$ and the mass of the CP-even Higgs boson, $m_h$.
 In particular, in the context of the mSUGRA-models, present
data on ${\cal B}(B \to X_s \gamma)$ \cite{cleobsg,alephbsg} and lower 
bounds on $m_h$ \cite{ALEPHmh}
do not allow the effective coefficient $\cse(m_B)$ to have a positive sign
\cite{goto99}. However,
relaxing the GUT mass constraints on the parameters of the scalar 
superpotential, large-$\tan \beta$ solutions exist, which are compatible
with all present experimental constraints and predict a range of 
$m_h$-values which are beyond the reach of LEP experiments 
\cite{goto99}. Interestingly, these large-$\tan\beta$-solutions in 
non-minimal SUGRA models do admit positive values for $\cse(m_B)$
which are compatible in magnitude with the CLEO bounds. In a more
general SUSY framework, the allowed parameter space for flavor-violating
transitions is much larger. Thus, in the MIA-approach \cite{LMSS99},
not only the sign of $\cse$ but also that of $C_{10}$ may have
either value. As different dilepton invariant mass regions in $B \to X_s 
\ell^+ \ell^-$, the coefficients $\cse(m_B)$, $\cne(m_B)$ and $\ct$ are 
weighted differently, a detailed knowledge of the invariant mass 
distribution and the FB-asymmetry \cite{amm91}, together with the decay 
rate $B \to X_s \gamma$, is completely sufficient to determine these 
effective coefficients \cite{agm95}.\footnote{Note that $\cse$, $\cn$
and $\ct$ are Wilson coefficients (numbers), 
but $\cne$ is a function of the dilepton invariant mass and encodes also
the information from the long-distance contribution. We assume that new 
physics leaves the long-distance part largely intact.} With obvious changes, 
these remarks apply to 
the exclusive decays $B \to (K,K^*) \ell^+ \ell^-$ as well with the
proviso that form factor dependence introduces an additional uncertainty,
which we  estimate in this paper. A relatively stable quantity is the
value of the dilepton invariant mass for which the FB-asymmetry becomes zero
in the SM. This has been discussed in the context of a number of 
phenomenological models for the form factors \cite{burdman}. We 
argue here that using the results of the 
large-energy expansion technique (called LEET) \cite{LEET}, the uncertainty
in the zero of the FB-asymmetry in the decays $B \to K^* \ell^+ \ell^-$
due to the form factors can be shown to be minimal.  This yields a strikingly
simple relation between the coefficients $\cse$ and $\cne$ which
we present in this paper.

  This paper is organized as follows: In section 2, we introduce the 
effective Hamiltonian formalism for semileptonic rare $B$-decays. Section
3 contains the definitions and derivations of the form factors in the
decays $B \to (K,K^*) \ell^+ \ell^-$ using the Light cone-QCD sum rule 
approach. In section 4, we display the decay distributions for the
invariant dilepton mass spectra for $B \to (K,K^*) \ell^+ \ell^-$ and the 
FB-asymmetry for $B \to K^* \ell^+
\ell^-$. Section 5 contains our numerical results for the
branching ratios and the FB-asymmetry in the SM, including comparison 
with the available data. Comparative studies in a number of variants of the
supersymmetric models are presented in section 6. Section 7 contains a
brief summary and some concluding remarks.

\section{\bf Effective Hamiltonian}
\setcounter{equation}{0}

At the quark level, the rare semileptonic decay 
$b \to s  \ell^+ \ell^-$ can be described in terms of the effective 
Hamiltonian obtained by integrating out the top quark and $W^\pm$ 
bosons: 
\begin{equation}
        \he = -4 \frac{G_F}{\sqrt{2}}  V_{t s}^\ast  V_{tb}  
              \sum_{i=1}^{10} \c_i(\mu)  \o_i(\mu) \; . 
        \label{eq:he}
\end{equation}
In this paper, we use the Wilson-coefficients $C_i$ calculated in the 
naive dimensional regularization (NDR) scheme \cite{effhamburas}.

The above Hamiltonian leads to the following free quark decay amplitude: 
\begin{eqnarray}
        \m(b\to s\ell^+\ell^-) & = & \frac{G_F \alpha}{\sqrt{2}  \pi} \, 
                V_{t s}^\ast V_{tb} \, \left\{
                  \cne  \left[ \bar{s}  \g_\mu  L  b \right] \, 
                          \left[ \lb  \g^\mu  \l \right]
                + \ct  \left[ \bar{s}  \g_\mu  L  b \right] \, 
                         \left[ \lb  \g^\mu  \g_5  \l \right]
                \right. \nonumber \\
        & & \; \; \; \; \; \; \; \; \; \; \; \; \; \; 
            \; \; \; \; \; \; \; \; \; \left. 
                - 2 \mbh  \cse  \left[ \bar{s}  i  \sigma_{\mu \nu}  
                        \frac{\qh^{\nu}}{\sh}  R  b \right] 
                        \left[ \lb  \g^\mu  \l \right]
                \right\} \; .
        \label{eq:m}
\end{eqnarray}
Here, $L/R \equiv {(1 \mp \g_5)}/2$, $s = q^2$, $q=p_{+} +p_{-}$ 
where $p_{\pm}$ are the four-momenta of 
the leptons, respectively. We put $m_s/m_b = 0$, but keep 
the leptons massive. The hat denotes normalization in terms of the 
$B$-meson mass, $m_B$, e.g. $\sh=s/m_B^2$, $\mbh=m_b/m_B$.
Here and in the remainder of this work we shall denote by $m_b \equiv m_b(\mu)$
the $\overline{\rm MS}$ mass evaluated at a scale $\mu$ and by $ 
m_{b,pole}$ the pole mass of the $b$-quark.
To NLO in perturbation theory, they are related by:
\begin{equation}
\label{eq:mbpole}
m_b(\mu)=m_{b pole}  \left[ 1- \frac{4}{3} \frac{\alpha_s(\mu)}{\pi}
\left\{1-\frac{3}{4} \ln(\frac{m_{b pole}^2}{\mu^2})\right\} \right] \; .
\end{equation}
Note that $\m(b \to s \ell^+ \ell^-)$, although a free quark decay 
amplitude, contains
certain long-distance effects from the matrix elements of 
four-quark operators, $\langle \ell^+ \ell^- s | {\cal O}_i 
 | b \rangle$, $1\leq i \leq 6$, which usually are absorbed into
a redefinition of the short-distance
Wilson-coefficients. To be specific, we define, for exclusive
decays\footnote{For {\em inclusive} decays one has in
addition to take into account the ${\cal{O}}(\alpha_s)$ virtual and 
bremsstrahlung 
corrections to the matrix element $\langle \ell^+ \ell^- s | {\cal O}_9 |
b \rangle$ as calculated in \cite{jezkuhn}. For {\em exclusive}
decays, one can define an effective coefficient by including only the
virtual corrections. We do not include any perturbative corrections to the
partonic matrix elements. 
However,  corresponding corrections are included in the  nonperturbative 
matrix element over mesons.}, the effective
coefficient of the operator ${\cal O}_9 = e^2/(16\pi^2)(\bar{s}  \g_\mu  
L  b) (\lb  \g^\mu  \l )$ as
\begin{equation}
\cne (\hat{s}) = C_9 + {Y} (\hat{s}) \; ,
\label{eqn:c9eff}
\end{equation}
where $Y(\sh)$ stands for the
above-mentioned matrix elements of the four-quark operators. 
A perturbative calculation yields \cite{burasmuenz,misiakE}:
\begin{eqnarray} 
        {Y}_{\rm pert} (\sh) & = & g(\mc,\sh)
                \left(3 \, C_1 + C_2 + 3 \, C_3
                + C_4 + 3 \, C_5 + C_6 \right)
\nonumber \\
        &- &  \frac{1}{2} g(1,\sh)
                \left( 4 \, C_3 + 4 \, C_4 + 3 \,
                C_5 + C_6 \right) 
         - \frac{1}{2} g(0,\sh) \left( C_3 +   
                3 \, C_4 \right) \nonumber \\
        &+ &     \frac{2}{9} \left( 3 \, C_3 + C_4 +
                3 \, C_5 + C_6 \right) \; . 
\label{eq:y}
\end{eqnarray}
We work in leading logarithmic (LLog) approximation with the values of $C_i$ 
given in Table \ref{wilson}. Formulae can be seen in \cite{burasmuenz}.
\begin{table}
        \begin{center}
        \begin{tabular}{|c|c|c|c|c|c|c|c|c|c|}
        \hline
        \multicolumn{1}{|c|}{ $C_1$}       & 
        \multicolumn{1}{|c|}{ $C_2$}       &
        \multicolumn{1}{|c|}{ $C_3$}       & 
        \multicolumn{1}{|c|}{ $C_4$}       &
        \multicolumn{1}{|c|}{ $C_5$}       & 
        \multicolumn{1}{|c|}{ $C_6$}       &
        \multicolumn{1}{|c|}{ $C_7^{\rm eff}$}       & 
        \multicolumn{1}{|c|}{ $C_9$}       &
                \multicolumn{1}{|c|}{$C_{10}$} &
 \multicolumn{1}{|c|}{ $C^{(0)}$ }     \\
        \hline 
        $-0.248$ & $+1.107$ & $+0.011$ & $-0.026$ & $+0.007$ & $-0.031$ &
   $-0.313$ &   $+4.344$ &    $-4.669$    & $+0.362$     \\
        \hline
        \end{tabular}
        \end{center}
\caption{ \it Values of the SM Wilson coefficients used in the numerical
          calculations, corresponding to the central values 
          of the parameters given in Table \protect\ref{parameters}.
Here, $C_7^{\rm eff} \equiv C_7 -C_5/3 -C_6$, and for $C_9$ we use the 
NDR scheme and $C^{(0)} \equiv 3 C_1 + C_2 + 3 C_3 + C_4 + 3 C_5 + C_6$.} 
\label{wilson}
\end{table}
For the decays $B \to X_s \ell^+ \ell^-$ (likewise, for $B \to
(K,K^*) \ell^+ \ell^-$), and with $\hat s$ far below the
$c\bar c$ threshold, perturbation theory, augmented by power corrections, is
expected to yield a reliable estimate. The power corrections in $1/m_c^2$
can not be calculated near the threshold $s=4 m_c^2$ and in the resonance
regions, as the heavy quark expansion breaks down \cite{bir98}. So, a 
complete profile of the FCNC semileptonic decays can not at present be
calculated from first principles. Several phenomenological prescriptions 
for incorporating the nonperturbative contributions to
$Y(\sh)$ exist in the literature \cite{amm91,ks96,lsw98}.
The resulting uncertainties on $\cne$ and various distributions in
the inclusive decays have been worked out in 
\cite{ah98-1,ah98-2,melikhov} to which we refer for detailed discussions. 
In the present paper we use the two parametrizations due
to Kr\"uger and Sehgal \cite{ks96} and Ali, Mannel and Morozumi \cite{amm91},
and interpret the difference in results for $\cne$
as an estimate of the theoretical uncertainty.
 Nonperturbative effects originate in particular from resonance
corrections to the perturbative quark-loops included in $Y_{\rm
  pert}(\sh)$. Light-quark loops are suppressed by small
Wilson-coefficients, so it is essentially only the charm-loop that
matters. Ref.~\cite{amm91} suggests to add the $c\bar c$
resonance-contributions from $J/\Psi, \Psi^\prime,\dots,\Psi^{(v)}$ 
to the perturbative result, with the former parametrized in the form of a
phenomenological Breit-Wigner Ansatz \cite{deshpande}. $Y$ is then
given by
\begin{equation}
Y_{\rm amm}(\sh) = Y_{\rm pert}(\sh) + \frac{3 \pi}{\alpha^2} C^{(0)}
         \sum_{V_i = \psi(1s),..., \psi(6s)} \kappa_i
      \frac{\Gamma(V_i \rightarrow \ell^+ \ell^-)\, m_{V_i}}{
      {m_{V_i}}^2 - \sh \, {m_B}^2 - i m_{V_i} \Gamma_{V_i}}
\end{equation}
with $C^{(0)} \equiv 3 C_1 + C_2 + 3 C_3 + C_4 + 3 C_5 + C_6$.
The phenomenological factors $\kappa_i$ correct for the factorization 
approximation which with $N_C=3$ (also called {\it naive factorization}
\cite{NS97}) gives a too small branching fraction for $B\to K^{(*)} 
V_i$. They can be fixed from 
\begin{eqnarray}
{\cal{B}}(B\to K^{(\ast)} V_i \to K^{(\ast)} \ell^+ \ell^-)=
{\cal{B}}(B\to K^{(\ast)} V_i) {\cal{B}}(V_i \to\ell^+ \ell^-) \; ,
\end{eqnarray}
where the right-hand side is given by data \cite{pdg98}.
While in the literature for {\it inclusive} $B \to X_s \ell^+ \ell^-$ 
decays, one comes across  
a universal $\kappa_i(X_s) \equiv \kappa_1(X_s)= 2.3$ \cite{ligeti}, 
we have evaluated the individual factors for the lowest two $c\bar{c}$
resonances, shown in Table \ref{tab:kappa}. 
\begin{table}[b]
        \begin{center}
        \begin{tabular}{|c|c|c|} 
 \hline
    \multicolumn{1}{|c|}{$\kappa$}
      & \multicolumn{1}{|c|}{$J/\Psi$}
      & \multicolumn{1}{|c|}{$\Psi^\prime$} \\
        \hline
  $K$           &2.70                   & 3.51 \\
  $K^\ast$      &1.65                   & 2.36 \\
        \hline
        \end{tabular}
        \end{center}
\caption{\it Fudge factors in $B\to K^{(\ast)} J/\Psi, \Psi^\prime \to  
K^{(\ast)} \ell^+ \ell^- $ decays calculated using the LCSR form factors.}
\label{tab:kappa}
\end{table}
In our numerical analysis we use for the higher resonances 
$\Psi^{(ii)},\dots,\Psi^{(v)}$ the average of $J/\Psi$ and $\Psi^\prime$.
We have averaged over charged and neutral $B$ mesons if data are available.
Concentrating on $J/\Psi$ only and assuming that the inclusive case $X_s$ is
saturated by $K$ and $K^*$, we get $\kappa_1(X_s)=1.9$.
Note that only the combination $|C^{(0)} \kappa_i|$ can be fixed from
the $J/\psi, \psi^\prime$-data. However, we treat the phase of the 
$\kappa_i$ as
fixed to the one in the factorization approach. This is substantiated by
data in which the Bauer-Stech-Wirbel parameters $a_1$ 
and $a_2$ 
\cite{BSW} are  consistently determined, with $a_1$ coming out close to 
its perturbation theory value and the sign of $a_2/a_1$ is the one given 
by the factorization approach \cite{bh95,NS97}.

 In the AMM-approach, it is tacitly
assumed that the extrapolation of the Breit-Wigner form away from the
resonances could be used to estimate these power corrections reliably.

 The 
KS-approach \cite{ks96}, on the other hand, bypasses the 
perturbative/non-perturbative dichotomy by 
using the measured cross-section $\sigma(e^+ e^-\to\,$hadrons) 
together with the assumption of quark-hadron duality for large $\sh$  to 
reconstruct $Y(\sh)$ from its imaginary part by a dispersion relation.
However, perturbative contributions in $\sigma(e^+
e^-\to\,$hadrons) and $ B \to X_s \ell^+ \ell^-$ are not identical.
In particular, the perturbative part of $Y(\sh)$ has genuine hard
contributions proportional to $m_b^2$, which can neither be ignored nor
taken care of by the quark-duality argument. The
issue in this approach remains as to how much of the genuine perturbative 
contribution 
in $ B \to X_s \ell^+ \ell^-$ arising from the $c\bar{c}$-continuum should be
kept and there is at present no
unique solution to this problem, as argued in \cite{lsw98} to which we refer 
for further 
discussion of this point. As stated earlier, we shall take the difference 
between the AAM-based and KS-based approaches for the long-distance 
contributions as a theoretical systematic error.

\section{Form factors from QCD sum rules on the light-cone}
\setcounter{equation}{0}

Exclusive decays $B\to (K,K^*) \ell^+ \ell^-$ are
described in terms of matrix elements of the quark operators in
Eq.~(\ref{eq:m}) over meson states, which can be parametrized in terms of form
factors.

Let us first define the form factors of the transition
involving the  pseudoscalar mesons $B\to K$. The
non-vanishing matrix elements are ($q=p_B-p$)
\begin{equation}\label{eq:ff1}
\langle K(p) | \bar s \gamma_\mu b | B(p_B)\rangle  =  f_+(s) \left\{
(p_B+p)_\mu - \frac{m_B^2-m_K^2}{s} \, q_\mu \right\} +
\frac{m_B^2-m_K^2}{s} \, f_0(s)\, q_\mu,
\end{equation}
and
\begin{eqnarray}
\langle K(p) | \bar s \sigma_{\mu\nu} q^\nu (1+\gamma_5) b | B(p_B)\rangle
& \equiv &  \langle K(p) | \bar s \sigma_{\mu\nu} q^\nu b |
B(p_B)\rangle\nonumber\\
& = & i\left\{ (p_B+p)_\mu s - q_\mu (m_B^2-m_K^2)\right\} \,
  \frac{f_T(s)}{m_B+m_K}.\label{eq:ff2}
\end{eqnarray}
For the vector meson $K^*$ with polarization vector $\epsilon_\mu$, we
can define the semileptonic form factors of the $V-A$ current by
\begin{eqnarray}
\langle K^*(p) | (V-A)_\mu | B(p_B)\rangle & = & -i \epsilon^*_\mu 
(m_B+m_{K^*}) A_1(s) + i (p_B+p)_\mu (\epsilon^* p_B)\,
\frac{A_2(s)}{m_B+m_{K^*}}\nonumber\\
\lefteqn{+ i q_\mu (\epsilon^* p_B) \,\frac{2m_{K^*}}{s}\,
\left(A_3(s)-A_0(s)\right) +
\epsilon_{\mu\nu\rho\sigma}\epsilon^{*\nu} p_B^\rho p^\sigma\,
\frac{2V(s)}{m_B+m_{K^*}}\,.}\hspace*{2cm}\label{eq:ff3}
\end{eqnarray}
Note the exact relations
\begin{eqnarray}
 A_3(s) & = & \frac{m_B+m_{K^*}}{2m_{K^*}}\, A_1(s) -
\frac{m_B-m_{K^*}}{2m_{K^*}}\, A_2(s),\nonumber\\
A_0(0) & = & A_3(0), \nonumber\\
\langle K^* |\partial_\mu A^\mu | B\rangle & = & 2 m_{K^*}
(\epsilon^* p_B) A_0(s).
 \label{eq:A30}
\end{eqnarray}
The second relation in (\ref{eq:A30}) ensures that there is no kinematical
singularity in the matrix element at $s=0$.
The decay $B\to K^*\ell^+\ell^-$ is described by the above semileptonic form
factors and the following penguin form factors:
\begin{eqnarray}
\langle {K^*} | \bar s \sigma_{\mu\nu} q^\nu (1+\gamma_5) b |
B(p_B)\rangle & = & i\epsilon_{\mu\nu\rho\sigma} \epsilon^{*\nu}
p_B^\rho p^\sigma \, 2 T_1(s)\nonumber\\
& & {} + T_2(s) \left\{ \epsilon^*_\mu
  (m_B^2-m_{{K^*}}^2) - (\epsilon^* p_B) \,(p_B+p)_\mu \right\}\nonumber\\
& & {} + T_3(s)
(\epsilon^* p_B) \left\{ q_\mu - \frac{s}{m_B^2-m_{{K^*}}^2}\, (p_B+p)_\mu
\right\}\label{eq:T}
\end{eqnarray}
with
\begin{equation}
 T_1(0)  =  T_2(0). \label{eq:T1T2}
\end{equation}
All signs are defined in such a way as to render the form factors positive.
The physical range in $s$ extends from $s_{\rm min} = 0$ to
$s_{\rm max} = (m_B-m_{K,K^*})^2$.

Lacking a complete solution of non-perturbative QCD, one has to rely
on certain approximate methods to calculate the above form
factors. In this paper, we choose to calculate them by the QCD sum rules
on the light-cone (LCSRs). The method of LCSRs was first
suggested for the study of
weak baryon decays in \cite{BBK} and later extended to heavy meson decays
in \cite{chernB}. It is a nonperturbative approach which combines ideas
of QCD sum rules \cite{shifman}
with the twist expansion characteristic for hard exclusive
processes in QCD \cite{exclusive} and makes explicit use of the large
energy of the final state meson at small values of the
momentum transfer to leptons $s$. In this respect, the LCSR
 approach is complementary to lattice calculations
\cite{flynn}, which are mainly restricted to form factors at small recoil
(large values of $s$) and at present require the scaling behavior found in 
the context of the LCSRs to extrapolate to smaller values of $s$
\cite{ABS94}. Of course, the LCSRs lack the
rigor of the lattice approach. Nevertheless, they prove to provide
a powerful nonperturbative model which is explicitly consistent with
perturbative QCD and the heavy quark limit.

Early studies of exclusive $B$ decays in the LCSR
approach were restricted to contributions of leading twist and
did not take into account radiative corrections. These corrections,
included in the estimates presented here, turn  out shift the form 
factors by $\sim 10$\%. 

\begin{table}
\addtolength{\arraycolsep}{3pt}
\renewcommand{\arraystretch}{1.4}
$$
\begin{array}{|l|lll|llll|lll|}
\hline
& f_+ & \phantom{-}f_0 & f_T & A_1 & A_2 & A_0 & V & T_1 & T_2 & T_3\\ \hline
F(0) & 0.319 & \phantom{-}0.319 & 0.355 & 0.337 & 0.282 & 0.471 &
0.457 & 0.379 & 0.379 & 0.260\\
c_1 & 1.465 & \phantom{-}0.633 & 1.478 & 0.602 & 1.172 & 1.505 & 1.482
& 1.519 & 0.517 & 1.129\\
c_2 & 0.372 & -0.095 & 0.373 & 0.258 & 0.567 & 0.710 & 1.015 & 1.030 &
0.426 & 1.128\\
c_3 & 0.782 & \phantom{-}0.591 & 0.700 & 0 & 0 & 0 & 0 & 0 & 0 & 0\\ 
\hline
\end{array}
$$
\caption[]{\it Central values of parameters for the parametrization 
  (\protect{\ref{eq:para}}) of
  the $B\to K$ and $B\to K^*$ form factors. Renormalization scale for
  the penguin form factors $f_T$ and $T_i$ is $\mu = m_b$. $c_3$ can
  be neglected for $B\to K^*$ form factors.}\label{tab:p1}
$$
\begin{array}{|l|lll|llll|lll|}
\hline
& f_+ & \phantom{-}f_0 & f_T & A_1 & A_2 & A_0 & V & T_1 & T_2 & T_3\\ \hline
F(0) & 0.371 & \phantom{-}0.371 & 0.423 & 0.385 & 0.320 & 0.698 &
0.548 & 0.437 & 0.437 & 0.295\\
c_1 & 1.412 & \phantom{-}0.579 & 1.413 & 0.557 & 1.083 & 1.945 & 1.462
& 1.498 & 0.495 & 1.044\\
c_2 & 0.261 & -0.240 & 0.247 & 0.068 & 0.393 & 0.314 & 0.953 & 0.976 &
0.402 & 1.378\\
c_3 & 0.822 & \phantom{-}0.774 & 0.742 & 0 & 0 & 0 & 0 & 0 & 0 & 0\\ 
\hline
\end{array}
$$
\caption[]{\it Parameters for the maximum allowed form 
factors.}\label{tab:p2} $$
\begin{array}{|l|lll|llll|lll|}
\hline
& f_+ & f_0 & f_T & A_1 & A_2 & A_0 & V & T_1 & T_2 & T_3\\ \hline
F(0) & 0.278 & 0.278 & 0.300 & 0.294 & 0.246 & 0.412 &
0.399 & 0.334 & 0.334 & 0.234\\
c_1 & 1.568 & 0.740 & 1.600 & 0.656 & 1.237 & 1.543 & 1.537
& 1.575 & 0.562 & 1.230\\
c_2 & 0.470 & 0.080 & 0.501 & 0.456 & 0.822 & 0.954 & 1.123 & 1.140 &
0.481 & 1.089\\
c_3 & 0.885 & 0.425 & 0.796 & 0 & 0 & 0 & 0 & 0 & 0 & 0\\ 
\hline
\end{array}
$$
\caption[]{\it Parameters for the minimum allowed form 
factors.}\label{tab:p3} \end{table}
\begin{figure}
\vskip 0.0truein
\centerline{\epsfysize=3.2in   
{\epsffile{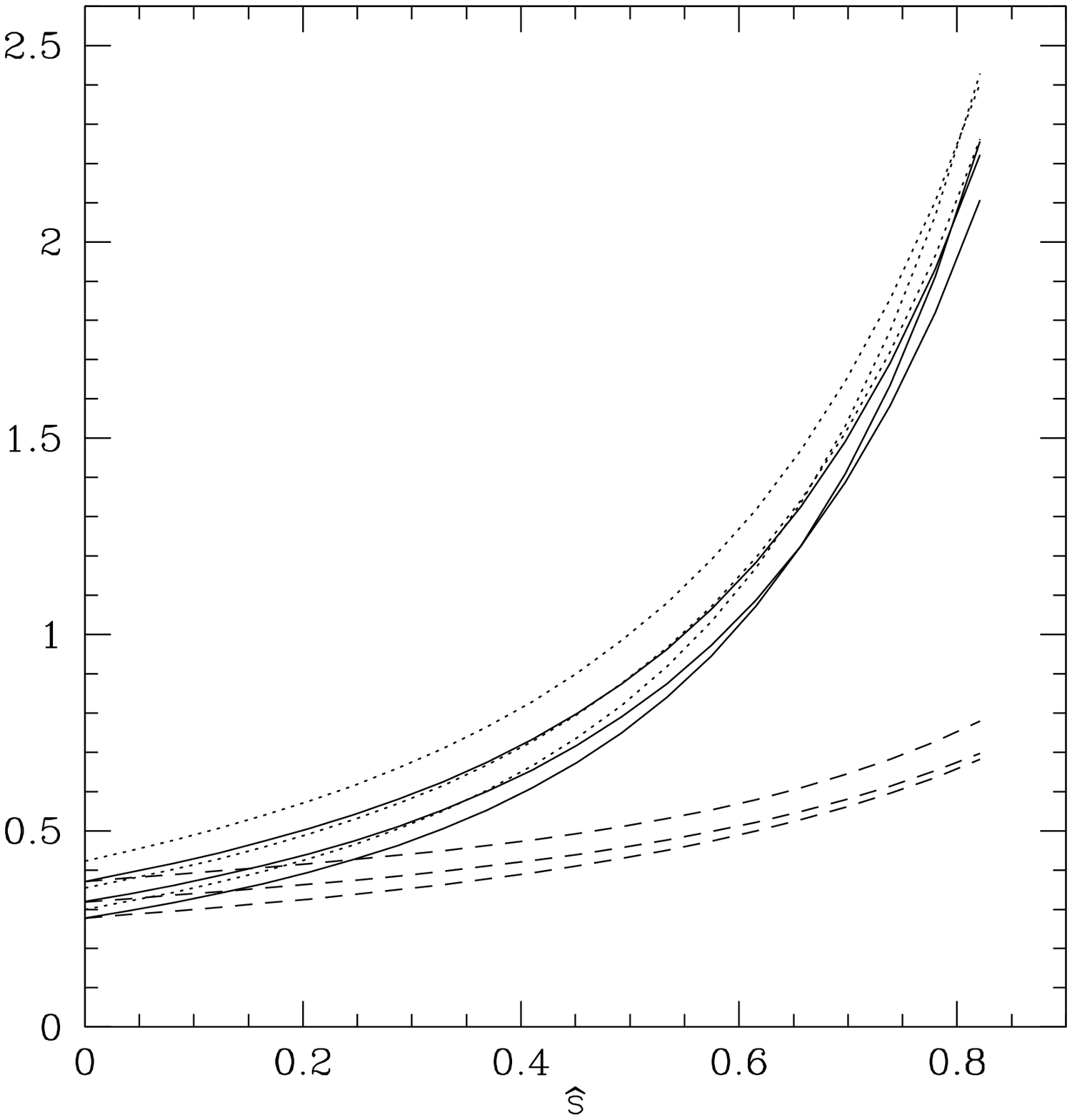}
}}
\vskip 0.0truein
\caption[]{ \it LCSR form factors with theoretical uncertainties 
for the $B \to K$ transition as a function of $\sh$.
Solid, dotted and dashed curves correspond to $f_+,f_T,f_{0}$,
respectively. Renormalization scale for $f_T$ is $\mu =
m_b$.}\label{fig:FFPseudo}
\vskip 20pt
     \begin{minipage}[t]{8.2cm}
     \mbox{ }\hfill\hspace{1cm}(a)\hfill\mbox{ }
     \epsfig{file=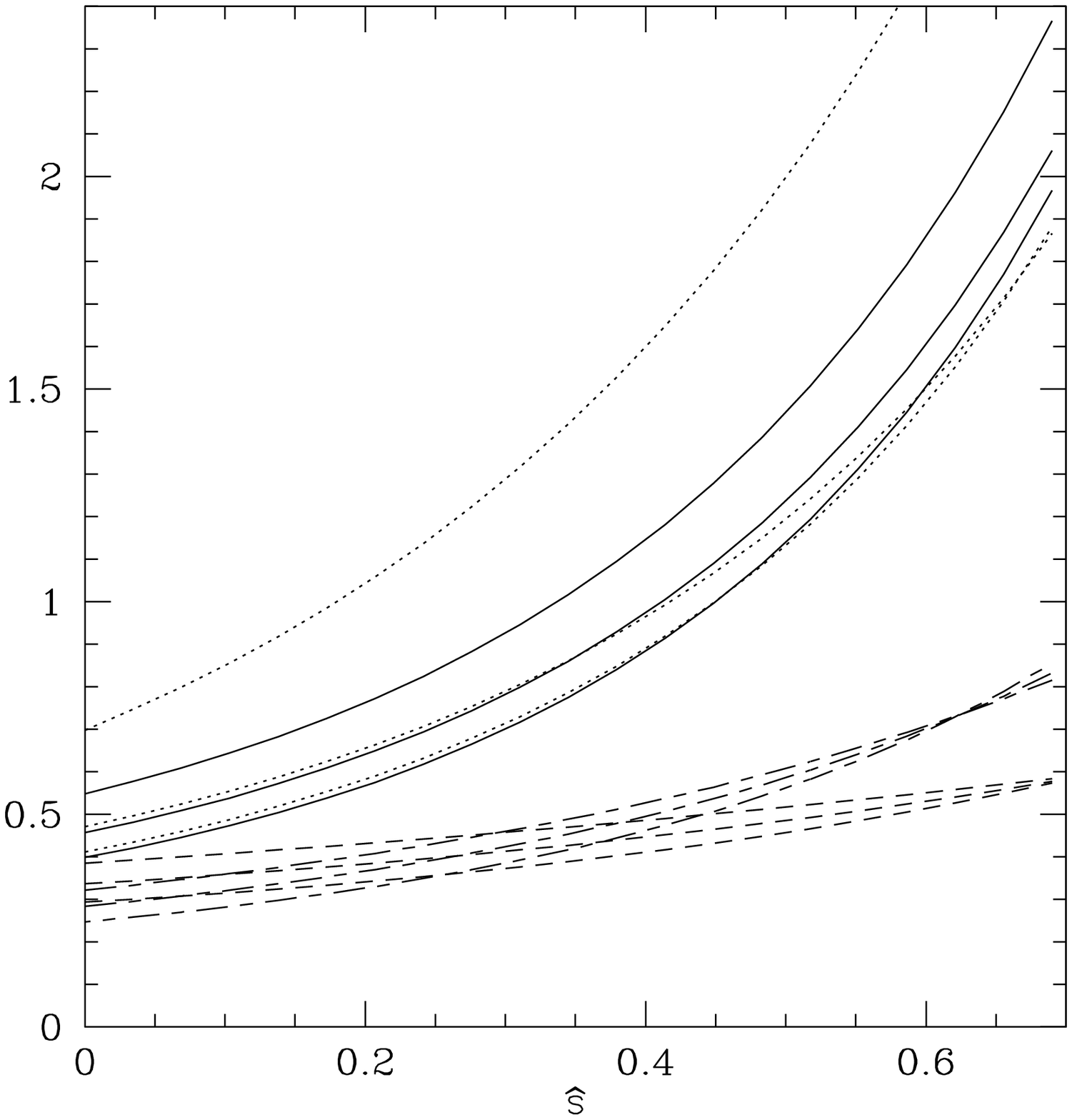,width=8.2cm}
     \end{minipage}
     \hspace{-0.4cm}
     \begin{minipage}[t]{8.2cm}
     \mbox{ }\hfill\hspace{1cm}(b)\hfill\mbox{ }
     \epsfig{file=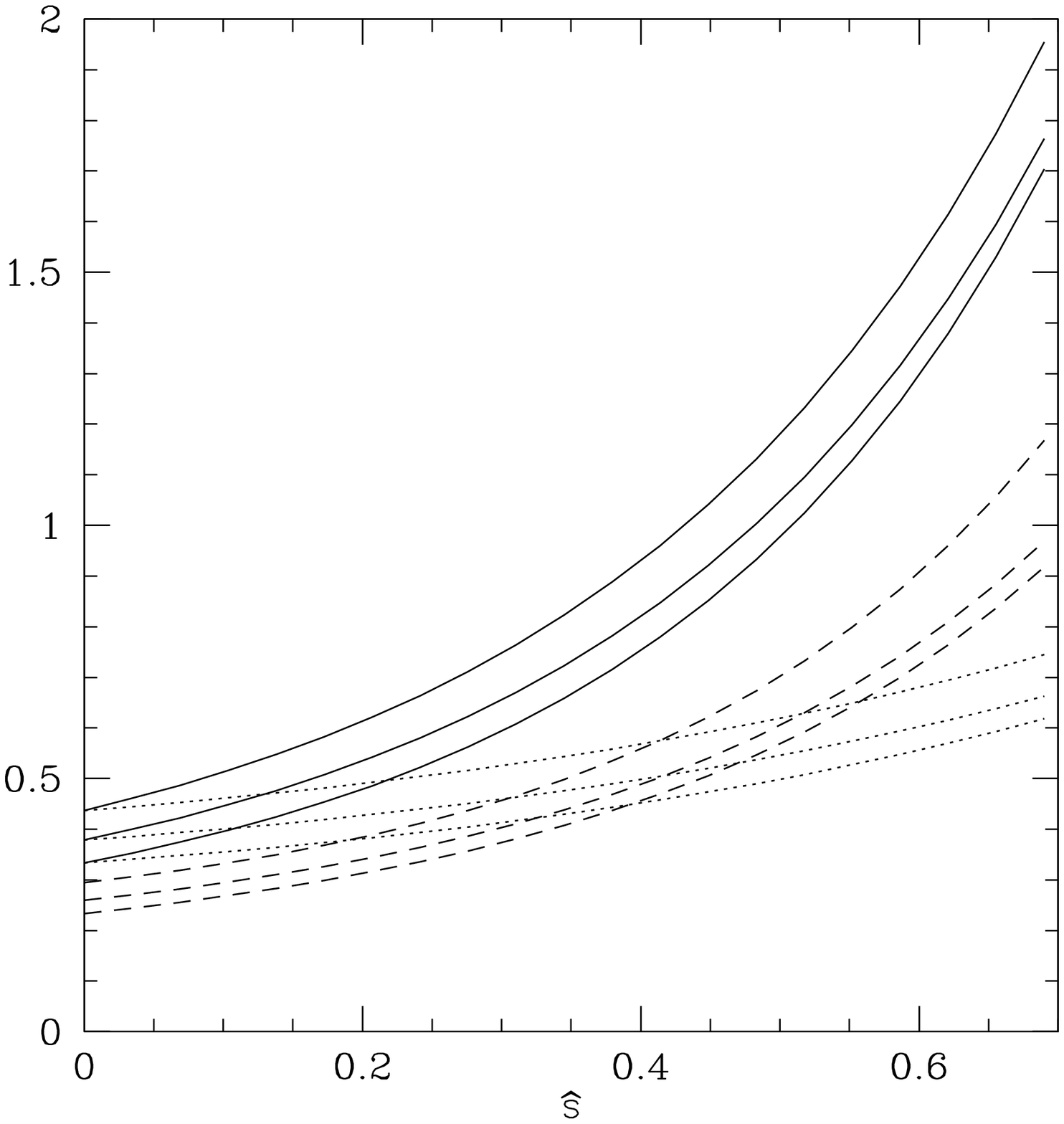,width=8.2cm}
     \end{minipage}  
     \caption{\it LCSR form factors with theoretical uncertainties for the 
$B \to K^\ast$ transition
as a function of $\sh$. In (a), the solid, dotted, dashed and
short long dashed curves correspond to $V,A_0,A_1,A_2$  and 
in (b), the solid, dotted and dashed  curves correspond to 
$T_1,T_2,T_3$, respectively. Renormalization scale for $T_i$ is $\mu =
m_b$.}
\label{fig:FFvector}
\end{figure}

In previous works \cite{bpiletter,ball98,ballbraun98}, the resulting form
factors have been parametrized by a modified single pole formula,
$$
F(\hat{s}) = \frac{F(0)}{1-a_F \hat{s} + b_F \hat{s}^2}\,,
$$
obtained from a fit to the LCSR result in the region $\sh<0.54$. The
extrapolation of this parametrization to maximum $\hat{s}$ is prone to
spurious singularities below the physical cut starting at
$s=m_{B_s^*}^2$. In the present work we thus choose a different
parametrization which avoids this problem:
\begin{equation}\label{eq:para}
F(\hat{s}) = F(0) \exp ( c_1 \hat{s} + c_2 \hat{s}^2 + c_3 \hat{s}^3).
\end{equation}
The term in $\hat{s}^3$ turns out to be important in $B\to K$
transitions, where $\hat{s}$ can be as large as 0.82, but can be
neglected for $B\to K^*$ with $\hat{s}< 0.69$. The parametrization
formula works within 1\% accuracy for $s<15\,$GeV$^2$. For an estimate of the
theoretical uncertainty of these form factors, we have varied the
input parameters of the LCSRs, i.e.\ the b quark mass, the
Gegenbauer-moments of the $K$ and $K^*$ distribution amplitudes and
the LCSR-specific Borel-parameters $M^2$ and continuum threshold $s_0$
within their respective allowed ranges specified in
\cite{ball98,ballbraun98} and obtain the three sets of form factors given in
Tabs.~\ref{tab:p1}--\ref{tab:p3}, which represent,
for each $\hat{s}$, the central value, maximum and minimum allowed
form factor, respectively. We plot the form factors
in Figs.~\ref{fig:FFPseudo} and \ref{fig:FFvector}.

Our value of $T_1(0)$ is consistent with the CLEO measurement of
${\cal{B}}(B\to K^{\ast} \gamma)_{exp}=
(4.2 \pm 0.8 \pm 0.6) \cdot 10^{-5}$ \cite{cleobkstar}. {}From the
formula for the decay rate,
\begin{equation}
\Gamma(B \to K^* \gamma)= \frac{G_F^2 \alpha |V_{ts}^* V_{tb}|^2}{32 \pi^4}
m_b^2 m_B^3 (1- m_{K*}^2/m_B^2)^3 |\cse|^2 |T_1(0)|^2 \; ,
\end{equation}
the central values of the parameters given in Table~\ref{parameters},
$T_1(0)=0.379$ and 
with $\tau_B = 1.61\,$ps we 
find ${\cal{B}}(B\to K^{\ast} \gamma)_{th} = 4.4\cdot 10^{-5}.$

\section{\bf Decay Distributions}
\setcounter{equation}{0}

In this section we define various decay distributions whose
phenomenological analysis will be performed in the next section.

Eq.~(\ref{eq:m}) can be written as 
\begin{equation}
        \m  =  \frac{G_F  \alpha}{2 \sqrt{2} \pi} \, 
                V_{ts}^\ast  V_{tb}  m_B \, \left[
                  \t_\mu^1 \, \left( \lb \, \g^\mu \, \l \right)
                + \t_\mu^2 \, 
                  \left( \lb \, \g^\mu \, \g_5 \, \l \right) \right] \; ,
        \label{eq:med}
\end{equation}
where for $B\to K\ell^+\ell^-$,
\begin{eqnarray}
  \t_\mu^1 & = & \ap(\sh) \, \ph_\mu + \bp(\sh) \, \qh_\mu \; , 
   \label{eq:t1bpll}\\
  \t_\mu^2 & = & \cp(\sh) \, \ph_\mu + \rp(\sh) \, \qh_\mu \; ,
   \label{eq:t2bpll}
\end{eqnarray}
and for $B\to K^*\ell^+\ell^-$,
\begin{eqnarray}
  \t_\mu^1 & = & 
    A(\sh) \, \epsilon_{\mu\rho\alpha\beta} \ep^\rho \pbh^\alpha
    \pvh^\beta 
    - i B(\sh) \, \ep_\mu 
    + i C(\sh) \, (\ep \cdot \hat{p}_B) \ph_\mu  
    + i D(\sh) \, (\ep \cdot \hat{p}_B) \qh_\mu \; , 
    \label{eq:t1bvll}\\
  \t_\mu^2 & = & 
  E(\sh) \, \epsilon_{\mu\rho\alpha\beta} \ep^\rho \pbh^\alpha \pvh^\beta 
  - i F(\sh) \, \ep_\mu 
    + i G(\sh) \, (\ep \cdot \hat{p}_B) \ph_\mu  
    + i H(\sh) \, (\ep \cdot \hat{p}_B) \qh_\mu \; , 
    \label{eq:t2bvll}
\end{eqnarray}
with $p \equiv p_B + p_{K,K^*}$. Note that, using the equation of
motion for lepton fields, the terms in $\hat{q}_\mu$
in ${\cal T}^1_\mu$ vanish and those in ${\cal T}^2_\mu$ become
suppressed by one power of the lepton mass. This effectively
eliminates the photon pole in $B'$ for $B\to K$. 

The auxiliary functions above are 
defined as 
\begin{eqnarray}
\label{eq:aux1}
  \ap(\sh) & = & \cne(\sh) \, f_+(\sh) 
         + \frac{2 \mbh}{1 + \mph} \cse f_T(\sh) \; , \\
  \bp(\sh) & = & \cne(\sh) \, f_-(\sh) 
         - \frac{2 \mbh}{\sh} (1 - \mph) \cse f_T(\sh) \; , \\
  \cp(\sh) & = & \ct \, f_+(\sh) \; , \\
  \rp(\sh) & = & \ct \, f_-(\sh) \; , \\
  A(\sh) & = & \frac{2}{1 + \mvh} \cne(\sh) V(\sh) 
         + \frac{4 \mbh}{\sh} \cse T_1(\sh) \; , \\ 
  B(\sh) & = & (1 + \mvh) \left[ \cne(\sh) A_1(\sh) 
         + \frac{2 \mbh}{\sh} (1 - \mvh) \cse T_2(\sh) \right] \; , \\
  C(\sh) & = & \frac{1}{1 - \mvh^2} \left[ 
         (1 - \mvh) \cne(\sh) A_2(\sh) 
         + 2 \mbh \cse \left( 
           T_3(\sh) + \frac{1 - \mvh^2}{\sh} T_2(\sh) \right) \right] \; , \\
  D(\sh) & = & \frac{1}{\sh} \left[ \cne(\sh) \left(
       (1 + \mvh) A_1(\sh) - (1 - \mvh) A_2(\sh) - 2 \mvh A_0(\sh) \right) 
      \right. 
      \nonumber \\
   & & \left. \; \; \; \; \; \; \; \; \; \; 
      - 2 \mbh \cse T_3(\sh) \right] \; , \\
  E(\sh) & = & \frac{2}{1 + \mvh} \ct V(\sh) \; , \\
  F(\sh) & = & (1 + \mvh) \ct A_1(\sh) \; , \\
  G(\sh) & = & \frac{1}{1 + \mvh} \ct A_2(\sh) \; , \\
  H(\sh) & = & \frac{1}{\sh} \ct \left[
       (1 + \mvh) A_1(\sh) - (1 - \mvh) A_2(\sh) - 2 \mvh A_0(\sh) \right] \; .
\label{eq:aux2}
\end{eqnarray}
Note that the inclusion of the full $s$-quark mass dependence into the
above formulae can
be done by substituting $m_b \to m_b+m_s$ into all terms proportional to
$\cse T_1$ and $\cse f_T$ and $m_b \to m_b-m_s$ in $\cse T_{2,3}$,
since 
$O_7 \sim \bar{s} \sigma_{\mu \nu} \left[ (m_b+m_s)+(m_b-m_s) \gamma_5 \right] 
q^\nu b$.

We choose the kinematic variables $(\sh,\uh)$ to be
\begin{eqnarray}
  \sh & = & \qh^2 = (\ph_+ + \ph_-)^2 \; , \\
  \uh & = & (\pbh - \ph_-)^2 - (\pbh - \ph_+)^2  \; 
\end{eqnarray}
which are bounded as 
\begin{eqnarray}
  (2 \mlh)^2 \leq & \sh & \leq (1 - \hat{m}_{K,K^*})^2  \; ,
  \label{eq:sbound}\\
  -\uh(\sh) \leq & \uh & \leq \uh(\sh) \; ,
  \label{eq:ubound}
\end{eqnarray}
with $\ml=m_{\ell}/m_B$ and
\begin{eqnarray}
  \uh(\sh) & =& \sqrt{\la (1-4 \frac{\mlh^2}{\sh})}  \; , \\
 \la & \equiv& \la(1,\hat{m}_{K,K*}^2,\sh) =
1+\hat{m}_{K,K*}^4+\sh^2-2 \sh-2 \hat{m}_{K,K*}^2(1+\sh) \; .
\end{eqnarray}
Note that the variable $\uh$ corresponds to $\theta$, the angle 
between the momentum of the $B$-meson and the positively charged lepton 
$\ell^+$  in 
the dilepton CMS frame, through the relation $\uh = -\uh(\sh) \cos \theta$
\cite{amm91}.
Keeping the lepton mass, we find the double differential 
decay widths $\gl^K$ and $\gl^{K^*}$ for the decays $B\to
K\ell^+\ell^-$ and $B\to K^*\ell^+\ell^-$, respectively, as
\begin{eqnarray}
  \frac{\d^2 \gl^K}{\d\sh \d\uh} & = & 
  \frac{G_F^2  \alpha^2  m_B^5}{2^{11}  \pi^5} 
      \left| V_{ts}^\ast  V_{tb} \right|^2  \nonumber \\
& & \times\left\{ 
(|\ap|^2 +|\cp|^2)
(\la -\uh^2 ) 
\right. \nonumber \\
&& + \left. |\cp|^2 4 \ml^2 (2+2 \mph^2-\sh)
+Re( \cp \ddp^{*}) 8 \ml^2 (1-\mph^2)
+|\ddp|^2 4 \ml^2 \sh \right\} \; ,    \label{eq:ddwbpll} \\
  \frac{\d^2 \gl^{K^*}}{\d\sh \d\uh} & = & 
  \frac{G_F^2 \, \alpha^2 \, m_B^5}{2^{11} \pi^5} 
      \left| V_{ts}^\ast \, V_{tb} \right|^2 
        \nonumber \\
  & &  \times\Bigg\{ 
  \frac{|A|^2}{4}  
   \left(\sh (\la + \uh^2) + 4 \mlh^2 \la  \right) 
  + \frac{|E|^2}{4} \left(\sh (\la + \uh^2) - 4 \mlh^2 \la  \right)   
        \Bigg.
        \nonumber \\
  & & \Bigg.
  + \frac{1}{4 \mvh^2} \left[ 
  |B|^2 \left( \la - \uh^2 + 8 \mvh^2 (\sh +2 \mlh^2 ) \right)
  + |F|^2 \left( \la - \uh^2 + 8 \mvh^2 (\sh -4 \mlh^2) \right) \right] 
        \Bigg.
        \nonumber \\
  & & \Bigg.
  - 2 \sh \uh \left[ {\rm Re}(BE^\ast) + {\rm Re}(AF^\ast) \right]
        \Bigg.
        \nonumber \\
  & & \Bigg.
  + \frac{\la}{4 \mvh^2} \left[ |C|^2 (\la - \uh^2) 
    + |G|^2 \left( \la - \uh^2 + 4 \mlh^2 (2 + 2 \mvh^2 - \sh) \right) \right]
        \Bigg.
        \nonumber \\
  & & \Bigg.
  - \frac{1}{2 \mvh^2} \left[ 
  {\rm Re}(BC^\ast) (1 - \mvh^2 - \sh)(\la - \uh^2) 
  \right. 
  \nonumber \\
  & & \left. \; \; \; \; \; \; \; \; \; \; \; \; 
  + {\rm Re}(FG^\ast) 
\left( (1 - \mvh^2 - \sh)(\la - \uh^2) + 4 \mlh^2 \la \right) \right]
        \Bigg.
        \nonumber \\
  & & \Bigg.
  - 2 \frac{\mlh^2}{\mvh^2} \la \left[ 
    {\rm Re}(FH^\ast)
    - {\rm Re}(GH^\ast) (1 - \mvh^2) \right] 
 + |H|^2 \frac{\mlh^2}{\mvh^2} \sh \la
  \Bigg\} \; .
   \label{eq:ddwbvll}
\end{eqnarray}

\subsection{\bf Dilepton mass spectrum}

We now give formulas for the dilepton invariant mass spectra. 
Integrating over $\uh$ in the kinematical region 
given in Eq.~(\ref{eq:ubound}) we find
\begin{eqnarray}
  \frac{\d \gl^K}{\d\sh} & = & 
  \frac{G_F^2  \alpha^2  m_B^5}{2^{10} \pi^5} 
      \left| V_{ts}^\ast  V_{tb} \right|^2  \uh(\sh)  \nonumber \\
& & \times\left\{ 
(|\ap|^2 +|\cp|^2) 
( \la- \frac{\uh(\sh)^2}{3} ) 
\right. \nonumber \\
& & + \left. |\cp|^2 4 \ml^2 (2+2 \mph^2-\sh)
+ Re( \cp \ddp^{*}) 8 \ml^2 (1-\mph^2)
+|\ddp|^2 4 \ml^2 \sh \right\} \; ,\label{eq:dwbpll}\\
  \frac{\d \gl^{K^*}}{\d\sh} & = & 
  \frac{G_F^2 \, \alpha^2 \, m_B^5}{2^{10} \pi^5} 
      \left| V_{ts}^\ast  V_{tb} \right|^2 \, \uh(\sh)
        \nonumber \\
  & & \times   
\Bigg\{ 
\frac{|A|^2}{3} \sh \la (1+2 \frac{\mlh^2}{\sh})
+|E|^2 \sh \frac{\uh(\sh)^2}{3}  
        \Bigg.
        \nonumber \\
  & & + \Bigg. \frac{1}{4 \mvh^2} \left[ 
|B|^2 (\la-\frac{\uh(\sh)^2}{3} + 8 \mvh^2 (\sh+ 2 \mlh^2) ) 
          + |F|^2 (\la -\frac{ \uh(\sh)^2}{3} + 8 \mvh^2 (\sh- 4 \mlh^2)) 
\right]
        \Bigg.
        \nonumber \\
  & & +\Bigg.
   \frac{\la }{4 \mvh^2} \left[ |C|^2 (\la - \frac{\uh(\sh)^2}{3}) 
 + |G|^2 \left(\la -\frac{\uh(\sh)^2}{3}+4 \mlh^2(2+2 \mvh^2-\sh) \right) 
\right]
        \Bigg.
        \nonumber \\
  & & - \Bigg.
   \frac{1}{2 \mvh^2}
\left[ {\rm Re}(BC^\ast) (\la -\frac{ \uh(\sh)^2}{3})(1 - \mvh^2 - \sh) 
\nonumber  \right. \Bigg.\\
& & + \left.  \Bigg.
       {\rm Re}(FG^\ast) ((\la -\frac{ \uh(\sh)^2}{3})(1 - \mvh^2 - \sh) + 
4 \mlh^2 \la) \right] 
        \Bigg.
        \nonumber \\
  & & - \Bigg.
 2 \frac{\mlh^2}{\mvh^2} \la  \left[ {\rm Re}(FH^\ast)-
 {\rm Re}(GH^\ast) (1-\mvh^2) \right] +\frac{\mlh^2}{\mvh^2} \sh \la |H|^2
  \Bigg\} \; .
   \label{eq:dwbvll}
\end{eqnarray}
Both distributions agree with the ones obtained in \cite{geng-kao}.
In the limit $m_{\ell}\to 0$ 
the form factors $f_0$ (or $f_{-}$) and $A_0$ do 
not contribute.
Furthermore, since $|\cse| \ll |\cne|, |\ct|$, 
the influence of $f_T,T_3$ on the distributions is subdominant.
That means that roughly  
$\d \gl^K/\d\sh \sim |f_+|^2$
for $\ell=e,\mu$ in the low $\sh$ region below the $J/\Psi$, 
with a $\sim -12 \%$ effect coming from $ \cse f_T$ terms.
For $B \to K^*$, the $b \to s \gamma$ transition is more important: for 
$s < 1 \mbox{GeV}^2$ the photon pole is the dominant
contribution, and it still contributes $\sim -30\,$\% around $s\approx
3\,$GeV$^2$. 

\subsection{\bf Forward-backward-asymmetry}

The differential forward-backward-asymmetry (FBA) is defined as \cite{amm91}
\begin{equation}
  \frac{\d \a_{\rm FB}}{\d \sh} = 
        -\int_0^{\uh(\sh)} \d\uh \frac{\d^2\gl}{\d\uh \d\sh}
              + \int_{-\uh(\sh)}^0 \d\uh \frac{\d^2\gl}{\d\uh \d\sh} \; .
  \label{eq:dfba}
\end{equation}
The FBA vanishes in $B\to K\ell^+\ell^-$ decays
as can be seen from Eq.~(\ref{eq:ddwbpll}), since 
there is no term containing $\uh$ with an odd power. 
For $B\to K^*\ell^+\ell^-$ decays it reads as follows
\begin{eqnarray}
  \frac{\d \a_{\rm FB}}{\d \sh}& =& 
\frac{G_F^2 \, \alpha^2 \, m_B^5}{2^{10} \pi^5} 
      \left| V_{ts}^\ast V_{tb} \right|^2 \, \sh \uh(\sh)^2
      \left[ {\rm Re}(BE^\ast) + {\rm Re}(AF^\ast) \right] \nonumber \\
&=&
  \frac{G_F^2 \, \alpha^2 \, m_B^5}{2^{8} \pi^5} 
      \left| V_{ts}^\ast  V_{tb} \right|^2 \, \sh \uh(\sh)^2 \nonumber \\
& & \times  \ct 
\left[  {\rm Re}(\cne) V A_1+ \frac{\mbh}{\sh} \cse (V T_2 (1-\mvh)+
A_1 T_1 (1+\mvh)) \right] \; .
  \label{eq:dfbabvllex}
\end{eqnarray}
The position of the zero $\sh_0$ is given by
\begin{eqnarray}
{\rm Re}(\cne(\sh_0)) =- \frac{\mbh}{\sh_0} \cse 
\left\{\frac{T_2(\sh_0)}{A_1(\sh_0)} (1-\mvh)+
\frac{T_1(\sh_0)}{V(\sh_0)} (1+\mvh)\right\} \; ,
\label{eq:fbzero}
\end{eqnarray}
which depends on the value of $m_b$, the ratio 
of the effective coefficients $\cse/{\rm Re}(\cne(\sh_0))$, and  
the ratio of the form factors shown above.
It is interesting to observe that in  the Large Energy Effective Theory
(LEET) \cite{LEET}, both ratios of the form factors appearing in 
Eq.~(\ref{eq:fbzero}) have essentially no hadronic uncertainty, i.e.\,, all 
dependence on
the intrinsically nonperturbative quantities cancels, and one has simply:
\begin{eqnarray} 
\frac{T_2}{A_1}&=& \frac{1+\mvh}{1+\mvh^2-\sh} 
\left(1-\frac{\sh}{1-\mvh^2}\right) \; ,
\nonumber\\
\frac{T_1}{V}  &=& \frac{1}{1+\mvh} \; .\label{eq:FBA}
\end{eqnarray}
With these relations, one has a particularly simple form for the equation
determining $\sh_0$, namely
\begin{equation}
{\rm Re}(\cne(\sh_0)) =- 2 \frac{\mbh}{\sh_0} \cse
\frac{1-\sh_0}{1+m_{K^*}^2 -\sh_0} \; .
\label{eq:fbzeroleet}
\end{equation}

Thus, the precision on the zero-point of the FB-asymmetry in $B \to K^* 
\ell^+ \ell^-$ is determined essentially
by the precision of the ratio of the effective coefficients and $m_b$, 
making it at par with the corresponding quantity in the inclusive
decays $B \to X_s \ell^+ \ell^-$, for which the zero-point is given by
the solution of the equation ${\rm Re}(\cne(\sh_0)) = -  
\frac{2}{\sh_0} \cse$. We find 
the insensitivity of $\sh_0$ to the decay form factors in $B \to K^* 
\ell^+ \ell^-$ a remarkable result, which  has also been discussed in
\cite{burdman}. However, the LEET-based
result in Eq.~(\ref{eq:FBA}) stands theoretically on more rigorous grounds
than the arguments based on scanning a number of form factor models.
With the coefficients given in Table~\ref{wilson} and $m_b=4.4$ GeV, we find
$\sh_0=0.10$ (i.e.\ $s_0=2.9\, \mbox{GeV}^2$) in the SM.
{}From Eq.~(\ref{eq:fbzero}) it follows that there is no zero below the 
$c \bar{c}$ reonances if both $C_9$ and $\cse$ have the same sign as
predicted in some beyond-the-SM models.

{}From the experimental point of view the normalized FB-asymmetry is
more useful, defined as
\begin{equation}
  \frac{\d \bar{\a}_{\rm FB}}{\d \sh} = 
\frac{\d \a_{\rm FB}}{\d \sh}/\frac{\d \gl}{\d\sh}
\end{equation}
which is equivalent to the energy asymmetry \cite{cmw,ahhm96}.
A slightly different definition is
\begin{equation}
  \frac{\d \a^{\prime}_{\rm FB}}{\d \sh} = 
\frac{\d \a_{\rm FB}}{\d \sh}/\Gamma
\end{equation}
whose integral gives the {\it global} energy asymmetry
$\a_{FB}^{\prime}=\a_{FB}/\Gamma$.

We summarize the characteristics of our observables:
\begin{itemize}
\item $\frac{d {\cal{B}}}{ds}(B\to K \ell^+ \ell^-)$ and 
$\frac{d {\cal{B}}}{ds}(B\to K^* \ell^+ \ell^-)$
get maximal for maximal $|\cse|,|\cn|,|\ct|$ and 
${\rm sign}(\cse Re(\cne)) =+1$. 
\item $\frac{d \bar{{\cal{A}}}_{FB}}{ds}(B\to K^* \ell^+ \ell^-)$
is proportional to $\ct$ and has a characteristic zero
(barring the trivial solution $\ct =0$, which we do not entertain here) if 
Eq.~(\ref{eq:fbzero}) is satisfied, which requires
\begin{equation}
{\rm sign}(\cse Re(\cne)) =-1~.
\label{eq:fbsign}
\end{equation}
\end{itemize}
The condition in Eq.~(\ref{eq:fbsign}) provides a discrimination between
the SM and models having new physics. For example, this condition is
satisfied in the SM and the SUGRA models with low-$\tan \beta$, in which 
case the actual position of $\sh_0$ would provide the further discriminant.
However, it turns out that the allowed parameter space of the SUGRA models
with large-$\tan \beta$ yield ${\rm sign}(\cse Re(\cne)) =+1$ \cite{goto99}, 
leading to the result that the FB-asymmetry in these models is 
parametrically different.
In particular, in all such cases, there is no zero of the FB-asymmetry.

\section{Branching Ratios and FB-Asymmetry in SM}
\begin{figure}[p]
\vskip 0.0truein
\centerline{\epsfysize=5.5in
{\epsffile{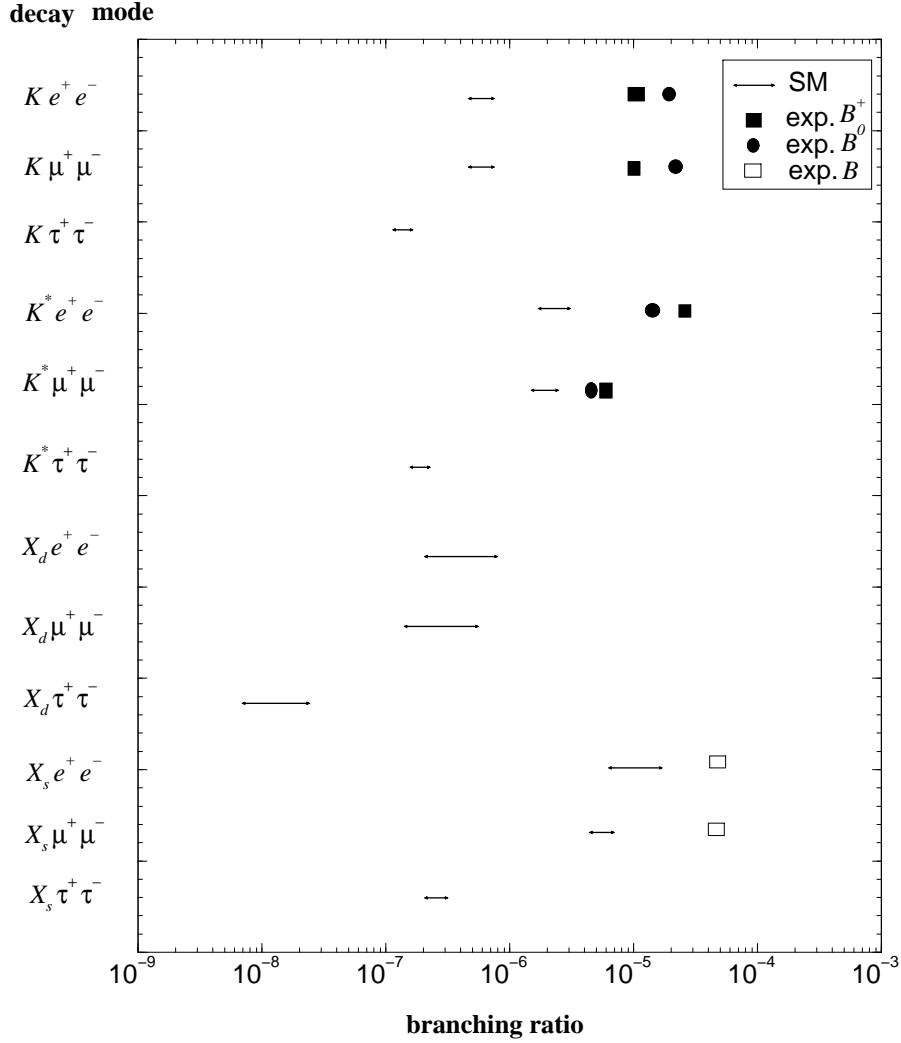}}}
\vskip 0.0truein
\caption[]{ \it Theoretical expectations for the exclusive decay 
branching ratios 
${\cal B}(B \to K^* \ell^+ \ell^-)$, ${\cal B}(B \to K \ell^+ \ell^-)$, 
$\ell^\pm =e^\pm,\mu^\pm,\tau^\pm$, calculated in
the LCSR method in the SM. For the sake of completeness, we also
give the branching ratios for the inclusive decays $B \to X_s \ell^+ \ell^-$
and $B \to X_d \ell^+ \ell^-$ in the SM, including the CKM dependence of the
latter.
Experimental upper limits (at 90\% C.L.) are also shown: solid squares
are from the charged $B^+$ decays (and charge conjugate),
 circles from the decays of $B^0$ (and charge conjugate),
and the empty squares are from the inclusive decays, averaged over the
charged and neutral $B$ decays. All experimental limits are from the CLEO 
\protect\cite{cleoexcl,cleobsll} and CDF \protect\cite{cdfexcl} 
collaborations.} \label{fig:smbrs}
\end{figure}

\setcounter{equation}{0}
The input parameters that we use in our numerical analysis are given in 
Table~\ref{parameters}. The parameters which are either well-known or 
have a small influence on the decay rates have been fixed to their central 
values, but we vary four of the listed parameters,
$m_t,~\mu,~m_{b,pole}$ and $\alpha_s(m_Z)$, in the indicated range. 
Furthermore, 
in the evaluation of the various distributions we use 
for $\mbh$ the MSbar mass evaluated at the scale $\mu=m_{b,pole}$, 
see Eq.~(\ref{eq:mbpole}).
%
In the SM we obtain the following non-resonant branching ratios, denoted by 
${\cal{B}}_{nr}$, ($\ell=e,\mu$):
\begin{eqnarray}
{\cal{B}}_{nr}(B \to K \ell^{+} \ell^{-})&=&5.7  \cdot 10^{-7}  , 
~~\Delta{\cal{B}}_{nr}= (^{+27}_{-15},\pm 6,^{+7}_{-6},\pm 1, \pm 
2)\%  , \label{eq:BKmu}\\
{\cal{B}}_{nr}(B \to K \tau^{+} \tau^{-})&=&1.3 \cdot 10^{-7}  , 
~~\Delta{\cal{B}}_{nr}=(^{+22}_{-6},\pm 
7,^{+4}_{-3},^{+0.4}_{-0.2}, \pm 1) \%  , \\
 {\cal{B}}_{nr}(B \to K^* e^{+} e^{-})&=&2.3 \cdot 10^{-6}  , 
~~\Delta{\cal{B}}_{nr}= (^{+29}_{-17}, ^{+2}_{-9},+ 12,^{+4}_{-1},\pm 
3)\%  , \\
 {\cal{B}}_{nr}(B \to K^* \mu^{+} \mu^{-})&=&1.9 \cdot 10^{-6}  , 
~~\Delta{\cal{B}}_{nr}=(^{+26}_{-17}, \pm 6, ^{+6}_{-4}, ^{-0.7}_{+0.4}, 
\pm 2)\%  , \label{eq:BKstmu}\\
{\cal{B}}_{nr}(B \to K^* \tau^{+} \tau^{-})&=&1.9 \cdot 10^{-7} , 
~~\Delta{\cal{B}}_{nr}=(^{+4}_{-8},\pm 
4,^{+13}_{-11},^{+0.6}_{-0.3},\pm 3) \%  .
 \end{eqnarray}
The first error in the $\Delta{\cal{B}}_{nr}$ consists of hadronic 
uncertainties from the form factors.
The other four errors given in the parentheses are due to 
the variations of $m_t,~\mu,~m_{b,pole}$ and $\alpha_s(m_Z)$, 
in order of appearance. In addition, there is an error of $\pm 2.5 \%$ from 
the lifetimes $\tau_B$ \cite{pdg98}.
The scale-dependence of the branching ratio ${\cal{B}}_{nr}(B \to K^* 
e^{+} e^{-})$ gives $+12\%$ and $+1.4\%$, as $\mu$ is varied from 
$\mu=m_{b,pole}$ to $\mu=m_{b,pole}/2$ and $\mu=2 m_{b,pole}$, respectively,
and we have taken the larger of the two errors in this case to estimate
the scale-dependence of this branching ratio.   
The largest parametric errors are
from the uncertainties of the scale $\mu$ and the top quark mass, $m_t$. 
The large 
scale-dependence of the branching ratios reflects essentially that of
the effective coefficients. To remedy this, one has to calculate the
virtual corrections to the matrix elements of the partonic decays
$b \to s \ell^+ \ell^-$  
to obtain perturbatively improved effective coefficients which are both 
scale- and scheme-independent \cite{AG97}. The exclusive decay form factors,
obtained in the LCSR method including the radiative corrections, depend
also on $m_b,\alpha_s$ and the 
renormalization scale $\mu$. However, the various dependencies of the form 
factors are inadequate to
compensate for the corresponding dependencies in the effective coefficients
being used.
\begin{table}
        \begin{center}
        \begin{tabular}{|l|l|}
        \hline
        $m_W$                   & $80.41$ GeV \\   
        $m_Z$                   & $91.1867$ GeV \\
        $\sin^2 \theta_W $      & $0.2233$ \\
        $m_c$                   & $1.4$ GeV \\
        $m_{b pole}$                   & $4.8 \pm 0.2 $ GeV \\
        $m_t$                   & $173.8 \pm 5.0$ GeV     \\
        $\mu$                & ${m_{b,pole}}^{+m_{b,pole}}_{-m_{b,pole}/2}$\\
        $\Lambda_{QCD}^{(5)}$   & $0.220^{+0.078}_{-0.063}$ GeV       \\
        $\alpha^{-1}$     & 129           \\
        $\alpha_s (m_Z) $       & $0.119 \pm 0.0058$ \\
        $|V^\ast_{ts} V_{tb}|$ & 0.0385 \\
        $|V^\ast_{ts} V_{tb}|/|V_{cb}| $ & 1 \\
        \hline
        \end{tabular}
        \end{center}
\caption{\it Default values of the input parameters and the $\pm 1~\sigma$
errors on the sensitive parameters used in our numerical calculations.}
\label{parameters}
\end{table}
 We present in Fig.~\ref{fig:smbrs} the exclusive branching ratios
calculated in the LCSR approach, obtained by adding the stated errors in
quadrature. We also give, for the sake of completeness, the branching 
ratios for the inclusive decays $B \to (X_s,X_d) \ell^+ \ell^-$.
In calculating the theoretical dispersion on $B \to X_d \ell^+ \ell^-$,
we have varied the CKM parameters in the allowed range obtained from the
CKM unitarity fits \cite{al99}. 
We have also listed the present experimental bounds
on the exclusive decays $B \to (K,K^*) e^+ e^-$ and
$B \to (K,K^*) \mu^+ \mu^-$, obtained by 
the CDF \cite{cdfexcl} and CLEO \cite{cleoexcl} collaborations. 
Experimental upper limits on the 
inclusive decays $B \to {X_s} e^+ e^-$ and $B \to {X_s} \mu^+ 
\mu^-$ are
from the CLEO collaboration \cite{cleobsll}. All experimental limits are
90\% C.L., and for the sake of this figure we have averaged the branching
ratios for the charged and neutral $B$-meson decays, as the differences in
their branching ratios are expected to be minimal theoretically.

  Figure \ref{fig:smbrs} shows that the exclusive decays $B \to 
K^* \mu^+ \mu^-$ and $B \to K^* e^+ e^-$ provide at present the most
stringent bounds on the effective coefficients. While none of the 
experimental bounds has reached the SM-sensitivity, they do provide
interesting upper limits on the parameter space of models with physics
beyond the SM.
 We will discuss this 
point in detail below in the context of the SUSY models we are studying 
in this paper.
We have also estimated the present theoretical precision on the quantity 
$s_0$ (zero of the FB-asymmetry) in the decay $B \to K^* \ell^+ \ell^-$
for $\ell^\pm =e^\pm$ and $\ell^\pm =\mu^\pm$. Note, that due to the
kinematics, there is no zero for the FB-asymmetry for the case
$B \to K^* \tau^+ \tau^-$.
Theoretical uncertainties from the form factors and the four parameters 
discussed above, $m_t,~\mu,~m_{b,pole}$ and $ \alpha_s(m_Z)$, on 
the $s_0$ are estimated as: 
$\pm 1\%, \pm 0.3\%,+14\%/-7\%,\pm 6\%,\pm 4\%$, respectively. As 
discussed above, the form factor-dependent uncertainty for this quantity 
is minimal, and the main sources of errors are $\mu$ and $m_{b,pole}$.
The reason of the marked scale dependence is again the lack of
compensating  perturbative 
corrections, in the absence of which the scale-dependence of the Wilson
coefficients reflects itself in rendering $s_0$ rather imprecise. 
Adding the stated errors in quadrature, we estimate in the SM
(fixing $m_b$ while varying $\mu$ and $\alpha_s(m_Z)$):
\begin{equation}
\label{eq:s0number}
s_0=2.88^{+0.44}_{-0.28}  \mbox{GeV}^2 \; .
\end{equation}
The actual dilepton mass distributions and the FB-asymmetry for the
decays of interest in the SM will be given later, together with the
corresponding estimates in some variants of SUSY.

\section{The decays $B \to (K,K^*) \ell^+ \ell^-$ in SUSY Models}
\setcounter{equation}{0}
First studies of rare $B$-decays $B \to X_s \gamma$ and $B \to X_s \ell^+ 
\ell^-$ in the context of MSSM were carried out in 
\cite{bertolini,agm95,cmw}.\footnote{There is a wrong sign in the
chargino and neutralino box matching condition in \cite{bertolini}.
This sign discrepancy between \cite{bertolini} and \cite{cmw}
has already been mentioned by the latter.
We are grateful to T.~Goto and F.~Kr\"uger for clarifying
this point.} Since then,  
these studies have been updated by taking into account progress in
theory and experiments.
 We employ the following models to study the rare 
$B \to K^{(*)} \ell^+ \ell^-$ decays: (i) Minimal supergravity (mSUGRA), 
(ii) Relaxed SUGRA (rSUGRA), obtained from mSUGRA by relaxing the universal 
scalar mass condition at the GUT scale \cite{hw97,goto96,goto99}, (iii) 
Minimal flavor violating 
supersymmetric model (MFV) (in the sense that the flavor violation is solely 
due to the standard CKM mechanism and resides in the charged current 
sector) \cite{MFVbsg}, and (iv) the Mass insertion approximation (MIA)
\cite{LMSS99}.
The last of these models serves as a generic supersymmetric extension of 
the SM having non-CKM flavor violations.
We do not consider models with broken R-parity and assume that there are no 
new phases from {\it new physics} beyond the SM, or, equivalently, that the
constraints from the electric dipole moments of the neutron and charged 
lepton and indirect constraints from the decay $B \to X_s \gamma$ as well
as other FCNC processes render these phases innocuous. This covers an important
part of the supersymmetric parameter space, but not all. The issue of 
supersymmetric phases having measurable consequences in CP-violations in
$B$ and $K$ decays and EDMs of the neutron and charged lepton is still
far from being settled. As we have not studied CP-asymmetries in the
decays $B \to (K,K^*) \ell^+ \ell^-$, the neglect of 
additional CP-phases is not crucial to the analysis of the decay rates being 
presented here. 

The strongest constraint on the MSSM parameter space is coming from data
on $B \to X_s \gamma$ \cite{cleobsg}, given in Eq.~(\ref{bsgamcleo}). 
In terms of the Wilson coefficients, this puts a bound
on the modulus of $\cse$, given in Eq.~(\ref{c7effbound}) in the NLO
approximation.
The SM-based estimate of $\cse$ in the NLO precision is well within this 
range, which then restricts the otherwise allowed parameter space in 
the supersymmetric models we are considering.
To be consistent with the precision of
other contributions in $B \to X_s \ell^+ \ell^-$, and for comparison
with the rates and distributions in the SM,
we work with $\cse(m_{b,pole})$ in the LLA accuracy. This yields the
bounds (at 95\% C.L.)
\begin{equation} 
0.249 \leq \vert C_7^{eff,LLA}(\mu=4.8~\mbox{GeV}) \vert \leq 0.374~.
\label{eq:c7lla}
\end{equation}
We remind at the outset that the theoretical uncertainties in the decay 
rates are estimated by us 
to be typically $\pm 35\%$. Hence SUSY-searches in $B \to (K,K^*) \ell^+ 
\ell^-$ will be unambiguous only for {\it drastic} SUSY effects.

\begin{figure}[p]
\vskip 0.0truein
\centerline{\epsfysize=3.5in
{\epsffile{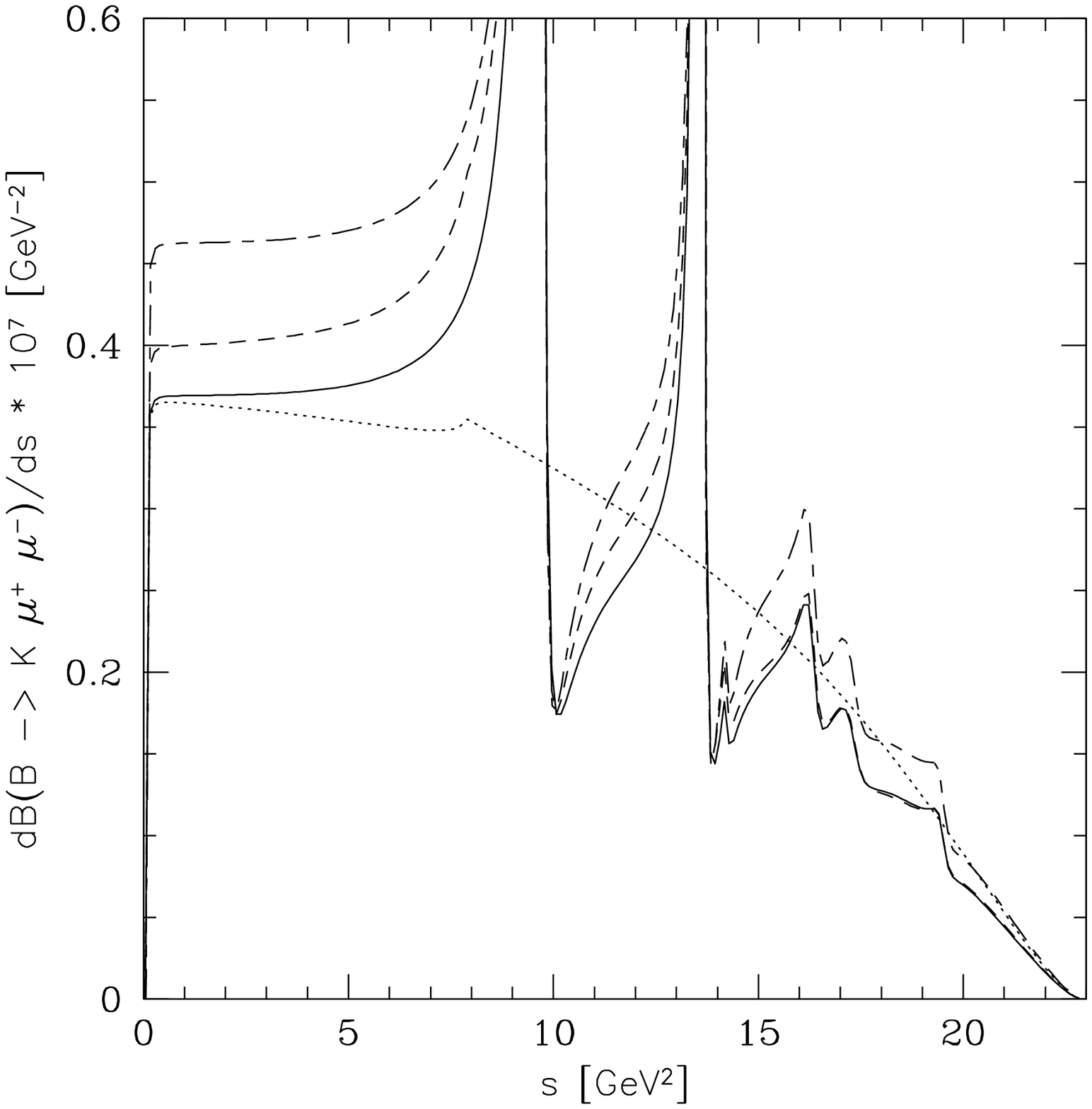}}}
\vskip 0.0truein
\caption[]{ \it The dilepton invariant mass distribution in
$B \to K \mu^+ \mu^-$ decays, using the form factors from LCSR
as a function of $s$. Solid curve: SD + LD using Ref.~\protect\cite{ks96},
dashed curve: SD + LD using Ref.~\protect\cite{amm91};
dotted: pure SD;
long-short dashed curve: SD + LD using Ref.~\protect\cite{ks96} with
$\cse=-\cse_{SM}$.}
 \label{fig:BK} 
%
%
%
%
     \mbox{ }\hspace{-0.7cm}
     \begin{minipage}[t]{8.2cm}
     \mbox{ }\hfill\hspace{1cm}(a)\hfill\mbox{ }
     \epsfig{file=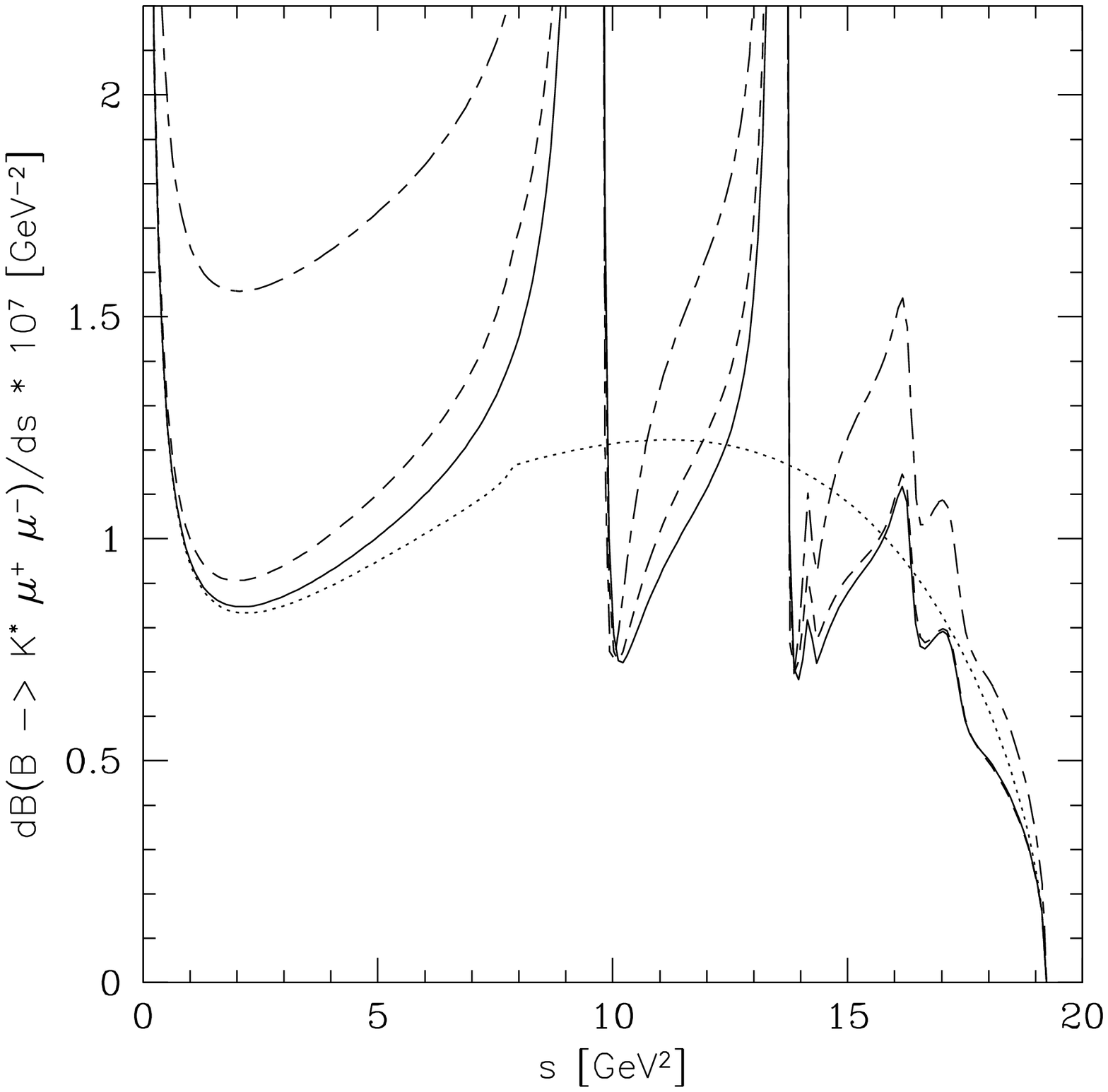,width=8.2cm}
     \end{minipage}
     \hspace{-0.4cm}
     \begin{minipage}[t]{8.2cm}
     \mbox{ }\hfill\hspace{1cm}(b)\hfill\mbox{ }
     \epsfig{file=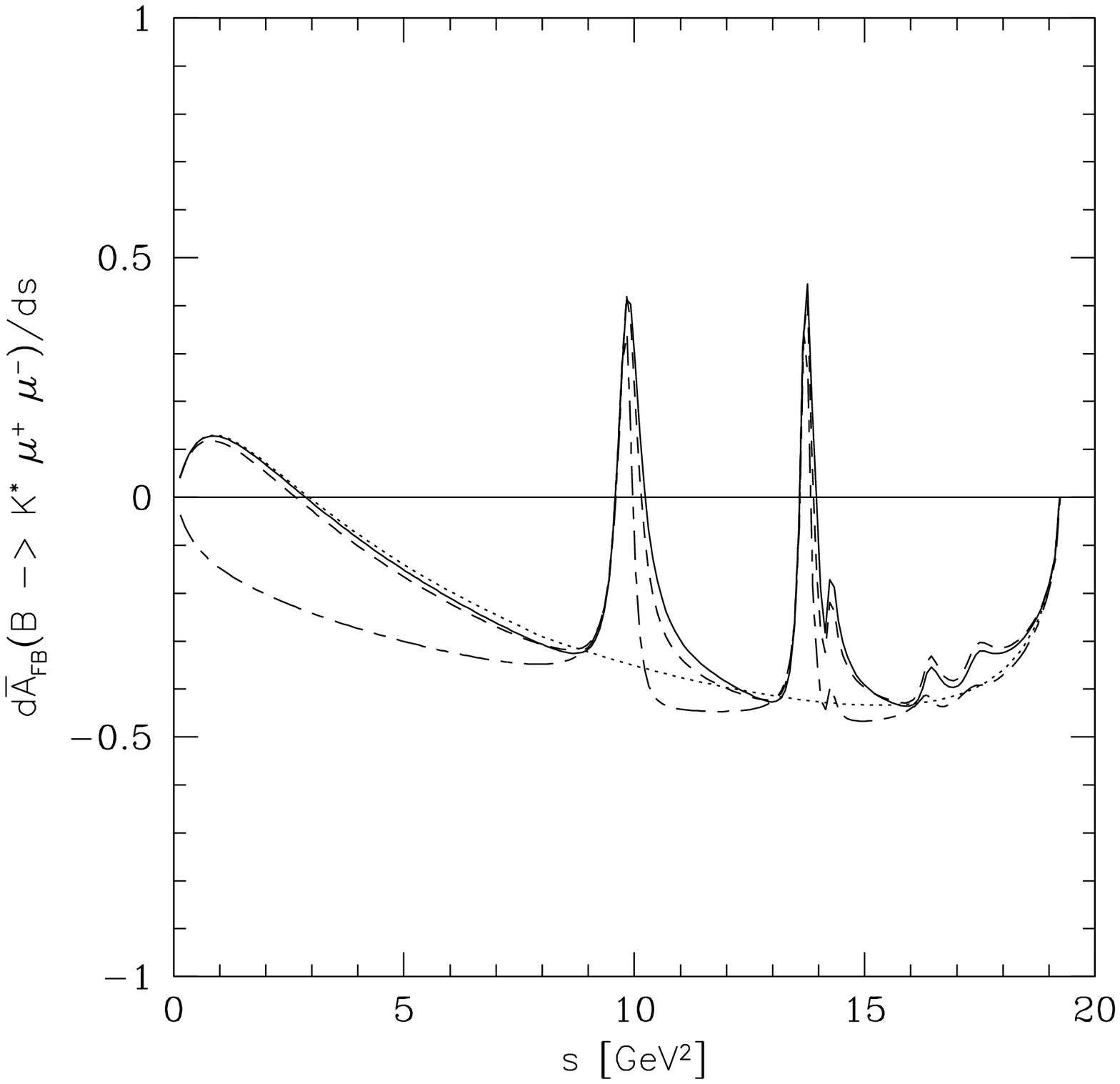,width=8.2cm}
     \end{minipage}
     \caption{\it The dilepton invariant mass distribution (a), and the
normalized FB-asymmetry (b) in
$B \to K^\ast \mu^+ \mu^-$ decays, using the form factors in LCSR
as a function of $s$ in the SM. Solid curves: SD + LD according to
Ref.~\protect\cite{ks96}, dashed curves: SD + LD using
Ref.~\protect\cite{amm91}; dotted: pure SD; 
long-short dashed curves: SD + LD using Ref.~\protect\cite{ks96} with
$\cse=-\cse_{SM}$.}
\label{fig:BKst}
\end{figure}

 To illustrate generic
SUSY effects in $B \to (K,K^*) \ell^+ \ell^-$, we start by assuming $|\cse| 
\simeq |\cse_{SM}|$ allowing 
for two possible solutions, $\cse <0$ (SM-like) and
$\cse >0$ (allowed in SUSY models). We also fix the other 
two coefficients $\cn$ and $\ct$ to their respective SM values. We show the
dilepton invariant mass distributions for 
$B \to K \mu^+ \mu^-$ and $B \to K^* \mu^+ \mu^-$ decays
in Figs.~\ref{fig:BK} and \ref{fig:BKst}(a), respectively. The FB-asymmetry
for $B \to K^* \mu^+ \mu^-$ is shown in Fig.~\ref{fig:BKst}(b).
These figures present a comparative study of the SM- and SUSY-based 
distributions, and the attendant 
theoretical uncertainties associated with the long-distance 
effects. For the latter, we have used the
KS-approach \cite{ks96} and the AMM-approach
\cite{amm91} to estimate the resonance-related uncertainties.  
These figures illustrate that despite non-perturbative uncertainties, it
will be possible to distinguish between the SM and a theoretical
scenario in which the magnitude of the effective coefficients are
similar, but  $\cse$ has the ''wrong sign". For the dilepton invariant 
mass, this reverses the sign of the interference term involving
${\rm Re}((\cse)^*\cdot\cne)$  which leads to significant difference 
in the decays $B \to K^* \ell^+ \ell^-$. More striking  
deviation from the SM prediction is found in $A_{FB}$
for the models in which the condition Eq.~(\ref{eq:fbsign}) is not
satisfied, resulting in a FB-asymmetry which remains negative below
the $J/\psi$-resonance region. This would be a {\it drastic deviation}
from the SM, which can not be fudged away due to non-perturbative
effects. Interestingly, the situation $\cse \simeq - \cse_{SM}$ is met
in a number of SUSY models as discussed below. In addition, in
a general flavor-violating supersymmetric model,  
also the other two Wilson coefficients ($\cn$ and $\ct$) may have either
sign. In this case, the FB-asymmetry in $B \to K^* \ell^+ \ell^-$ may
have a functional dependence on the dilepton mass which is 
characteristically different than the ones obtaining in the SM and SUGRA 
models, as shown below.

More elaborate changes from new physics (NP) in the values of the  
relevant Wilson coefficients can be taken into account by the (correlated) 
ratios, ($i=7,9,10$):
\begin{equation}
R_i(\mu)\equiv \frac{{\cal{C}}^{NP}_i+{\cal{C}}^{SM}_i}
{{\cal{C}}^{SM}_i}=\frac{{\cal{C}}_i}{{\cal{C}}^{SM}_i} \; ,
\end{equation}
which depend on the renormalization scale (except for $\ct$),
for which we shall always take $\mu=m_{b,pole}$. The
experimental constraint from $B \to X_s \gamma$ given in
Eq.~(\ref{bsgamcleo}) now translates into the bound
\begin{equation}  
0.80 < |R_7(\mu=4.8~\mbox{GeV})| < 1.20~,
\label{eq:R7bounds}
\end{equation}
where the coefficients are understood to be calculated in the LLA precision.
In the numerical estimates, we have used ${\cal B}_{sl}=(10.4\pm 0.4)\%$
for the average semileptonic branching ratio, and have set the heavy
quark expansion parameters to the values $\lambda_1=-0.20$ GeV$^2$
and $\lambda_2=0.12$ GeV$^2$. 
The allowed values of the other two ratios $R_{9}$ and $R_{10}$
are taken from the literature for the mSUGRA and rSUGRA models
\cite{goto96,goto99}, and for the other two models, MFV and MIA, we
have calculated them. In particular, in the MIA approach, large 
enhancements are anticipated in the branching ratio ${\cal B}(B \to X_s 
\ell^+ \ell^-)$ in some allowed region of the parameter space \cite{LMSS99}.
These enhancements, suitably modified by the form factors, are also 
present in the branching 
ratios for the exclusive decays $B \to (K,K^*) \ell^+ \ell^-$.  However,
as shown in Fig.~\ref{fig:smbrs}, some of these branching ratios are
bounded quite stringently, in particular, for the 
decays $B \to K^* e^+ e^-$ and $B \to K^* \mu^+ 
\mu^-$\cite{cdfexcl,cleoexcl}. Assuming $R_7$ in the allowed range, we shall
work out the constraints on the effective coefficients $C_{9}$ and $C_{10}$
(equivalently $R_9$ and $R_{10}$). Based on this analysis, we shall show
the dilepton invariant mass spectra and the FB-asymmetry in  
some representative cases.  
\subsection{$B \to (K,K^*) \ell^+ \ell^-$ in SUGRA models}
\begin{figure}[p]
\centerline{\epsfysize=3.3in   
{\epsffile{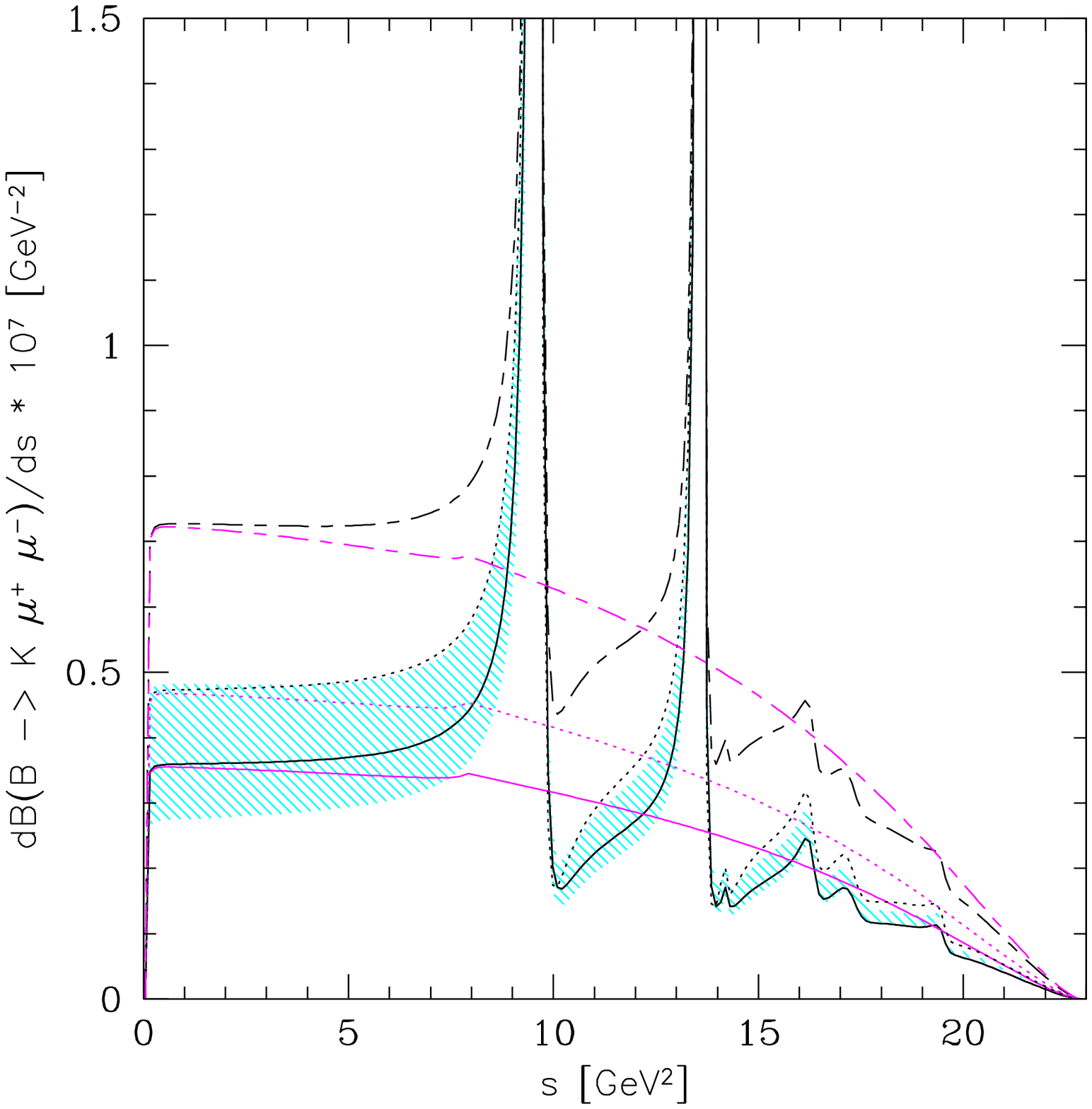}}}
\caption[]{ \it The dilepton invariant mass distribution in
$B \to K \mu^+ \mu^-$ decays, using the form factors from LCSR as a function 
of $s$. All resonant $c \bar{c}$ states are parametrized as 
in Ref.~\protect\cite{ks96}.
 The solid line represents the SM and the shaded area 
depicts the form factor-related uncertainties. 
The dotted line corresponds to the SUGRA model with
$R_7=-1.2,~R_9=1.03$ and $R_{10}=1$. The long-short dashed 
lines correspond to an allowed point in the parameter space of the 
MIA-SUSY model, given by $R_7=-0.83$, $R_9=0.92$ and $R_{10}=1.61$.  
The corresponding pure SD spectra are shown in the lower part of the plot.
}\label{fig:BKsusy}
 %
$$
\epsfysize=3.3in   
{\epsffile{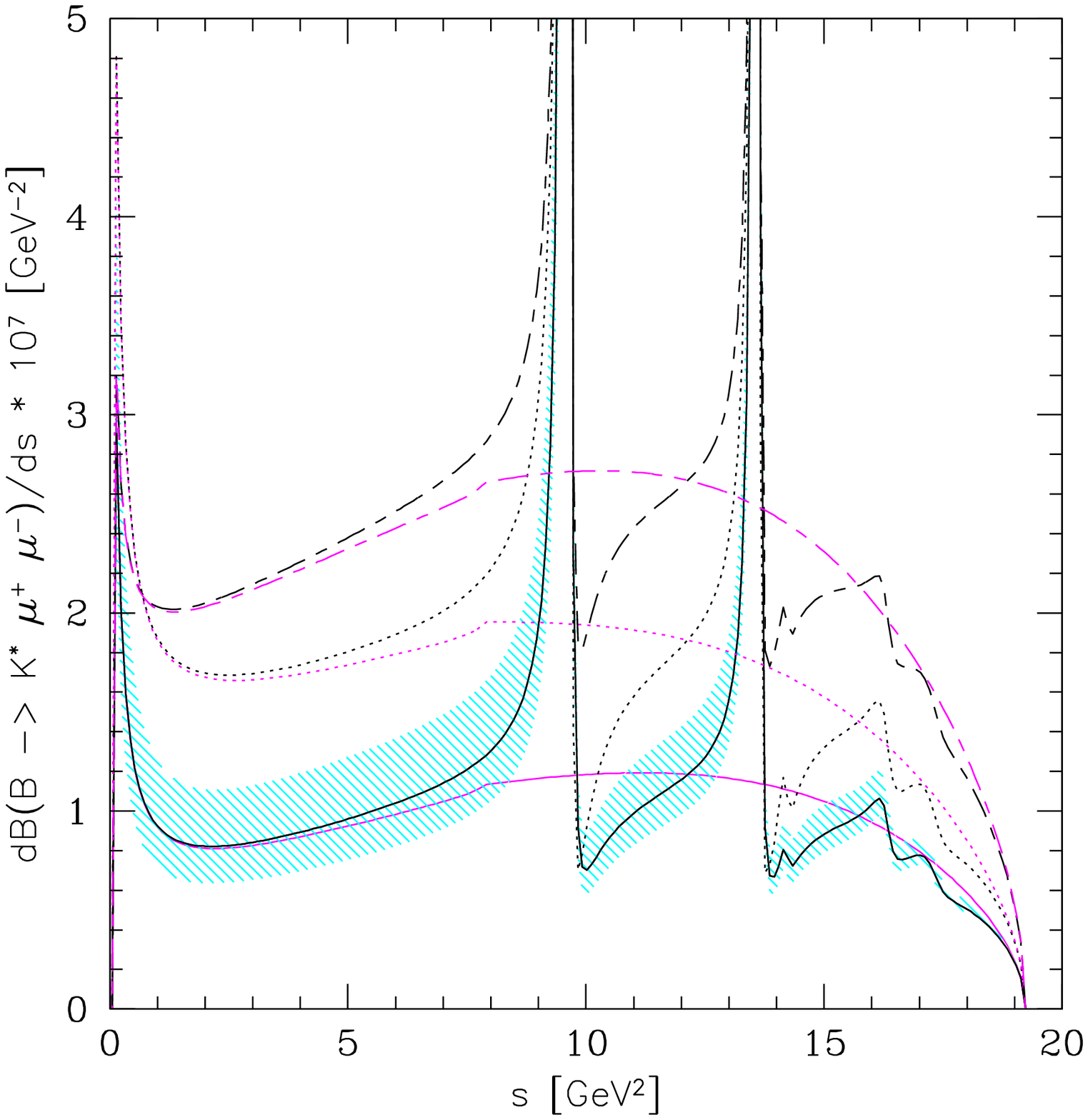}}
$$
\vspace*{-30pt}
\caption[]{ \it The dilepton invariant mass distribution in
$B \to K^* \mu^+ \mu^-$ decays, using the form factors from LCSR as a function
of $s$. All resonant $c \bar{c}$ states are parametrized as
in Ref.~\protect\cite{ks96}. The legends are the same as in 
Fig.~\protect\ref{fig:BKsusy}.}
\label{fig:BKstsusy}
\end{figure}

\begin{figure}[t]
\vskip 0.0truein
\centerline{\epsfysize=3.5in   
{\epsffile{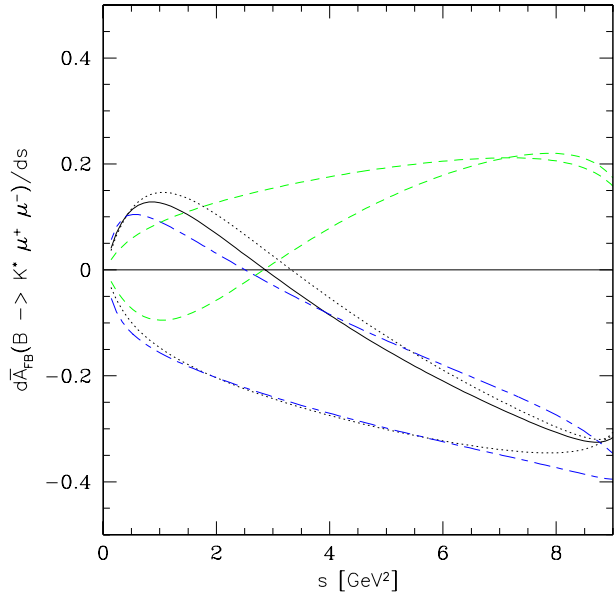}}}
\vskip 0.0truein
\caption[]{ \it The normalized forward-backward asymmetry in
$B \to K^* \mu^+ \mu^-$ decay as a function of $s$, using the form 
factors from the LCSR approach.
All resonant $c \bar{c}$ states are parametrized as
in Ref.~\protect\cite{ks96}. The
solid line denotes the SM prediction. 
The dotted (long-short dashed) lines correspond to the SUGRA  
(the MIA-SUSY) model, using the parameters given in 
Eq.~(\protect\ref{eq:sugrar}) (Eq.~(\protect\ref{eq:MIAplots})) with the 
upper and  lower curves representing the $\cse <0$ and $\cse >0$ case, 
respectively. The dashed curves indicating a positive asymmetry for
large $s$ correspond to the  MIA-SUSY models using the
parameters given in Eq.~(\ref{eq:bestdepr}), i.e.\  the ''best 
depression scenario" with $\ct>0$.}
\label{fig:BKstAFBsusy}
\end{figure}

 We shall consider here both the minimal and restricted 
SUGRA models (mSUGRA, rSUGRA). 
The parameter space of these models may be decomposed into two qualitatively 
different regions, which can be characterized by $\tan\beta$ values.
 For small $\tan\beta$, say $\tan\beta \sim 2$, the sign of 
$\cse$ is the same as in the SM. Here, no spectacular deviations from the 
SM can be expected in the decays $B \to (K,K^*) \ell^+ \ell^-$. Given the 
theoretical uncertainties shown earlier by us, we think that it would
be very difficult to disentangle any SUSY effects for this scenario in these 
decays.
For large $\tan\beta$, the situation is more interesting due to correlations
involving the branching ratio for $B \to X_s \gamma$, the mass of the 
lightest CP-even Higgs boson, $m_h$, and sign$(\mu_{susy})$, appearing in 
the Higgs superpotential. In this case,
 there are two branches for the solutions for $m_h$
and ${\cal B}(B \to X_s \gamma)$. The interesting scenario for
SUSY searches in $B \to (K,K^{*}) \ell^+ \ell^-$ is the one in which 
sign$(\mu_{susy})$ and $m_h$ admit $\cse$ to be positive. For example, 
this happens for $\tan\beta \geq 10$, in which case $m_h=(115$--$125)$ GeV and 
$\cse$ is positive and obeys the $B \to X_s \gamma$ bounds 
\cite{goto99}. Following the generic case shown earlier, one expects 
a constructive interference of the terms depending on $\cse$ and $\cn$ in 
the dilepton invariant mass spectra. For the sake of illustration, we 
use
\begin{equation}
R_7=-1.2,~~R_9=1.03,~~R_{10}=1.0 ~,
\label{eq:sugrar}
\end{equation}
obtained for 
$\tan\beta= 30$ \cite{goto96}, as a representative large-$\tan\beta$
solution, to study the effects on 
our observables. We find that in the low-$q^2$ region the branching ratio  
for $B\to K 
\mu^+ \mu^-$ is enhanced by about $30 \% $ compared to the SM one, as 
shown in Fig.~\ref{fig:BKsusy}. This enhancement is difficult to disentangle 
from the non-perturbative uncertainties attendant with the SM-distributions
(shown as the shaded band in this figure).
The dilepton mass distribution  
for $B \to K^* \mu^+ \mu^-$ is more promising, as in this case the 
enhancement is around $100 \%$, see Fig.~\ref{fig:BKstsusy}, and 
this is distinguishable 
from the SM-related theoretical uncertainties (shown as the shaded band in 
this figure). Note that the resulting branching ratios are consistent
with the present experimental upper bounds on these decays given earlier.
The supersymmetric effects presented here are  
very similar to the ones worked out for the inclusive decays
$B \to X_s \ell^+ \ell^-$ \cite{goto96}, where enhancements of
($50$--$100$)\% were predicted in the low-$q^2$ branching ratios. 
The effect of $R_7$ being negative is striking in the FB 
asymmetry as shown in Fig.~\ref{fig:BKstAFBsusy}, in which the two SUGRA 
curves are plotted using Eq.~(\ref{eq:sugrar}) (for $R_7 < 0$) and
by flipping the sign of $R_7$ but keeping the magnitudes
of $R_i$ to their values given in this equation.
Summarizing for the SUGRA theories, large $\tan\beta$ solutions lead to
$\cse$ being positive, which implies that    
FB-asymmetry below the $J/\psi$-resonant region remains negative (hence, no
zero in the FB-asymmetry in this region) and one expects an 
enhancement up to a factor two in the dilepton mass distribution in $B \to 
K^* e^+ e^-$  and $B \to K^* \mu^+ \mu^-$.

\subsection{$B \to (K,K^*) \ell^+ \ell^-$ in MFV-SUSY Model}
 The MFV-SUSY  model is based on the assumption of minimal 
flavor violation. Here, quarks and squarks are aligned so there is no  
flavor-changing $q - \tilde{q}^{\prime}-(\tilde{Z},\tilde{\gamma},\tilde{g})$ 
vertex and the charged
one, $d-\tilde{u}-\tilde{\chi}^\pm$, is governed by the CKM matrix.
As a consequence, in this model neutralino-down-squark and gluino-down-squark
graphs do not contribute to either $b \to s \gamma$ or $b \to s \ell^+ 
\ell^-$ transitions.  
In addition to the charged Higgs-top graphs, chargino-up type squarks loops 
with 
a light stop $\tilde{t}_1$, and the $W^\pm$-top quark loops, present in
the SM, give 
the dominant contribution. While not holding 
generally, the assumptions in the MFV-SUSY model are valid over an important
part of the minimal supersymmetric parameter space \cite{MFVbsg}. They have 
the simplifying
feature that the dominant supersymmetric effects remain confined to charged 
current transitions and relatively easy to test experimentally due to
well-defined correlations in several measurable quantities 
involving FCNC transitions \cite{MFVbsg,al99}.

As is well-known \cite{gbhiggs}, 
in the two-Higgs doublet model of type II (2HDM model II), which is
embedded also in the MFV-SUSY construct, the charged-Higgs  contribution is 
always additive to the SM, i.e., $\cse({\rm 2HDM}) <0$, yielding a 
lower bound on the charged Higgs mass $m_{H^{\pm}}$ (almost)
independent of $\tan\beta$, above $\tan \beta > 1$.
In MFV, the $\tilde{\chi}^\pm-\tilde{t}_1$ loop can compensate the $H^\pm-t$
contribution, with a large positive contribution to $\cse$.
We scan over the parameter space in the range 
$55 ~\mbox{GeV} < m_{H^\pm}< 1$ TeV, $0 < M_2, |\mu_{susy}| < 500$ GeV, 
where $\mu_{susy}$ is the bilinear Higgs coupling in the superpotential and 
$M_2$ is the gaugino soft breaking mass. 
We use $m_{\tilde{q}}=m_{\tilde{t}_2}=1$ TeV, where $m_{\tilde{q}}$ 
denotes the (degenerate) masses of
other than top squarks,  
and fix $m_{\tilde{\nu}}= 50$ GeV to its lower bound.
We reject too light charginos, demanding $m_{\tilde{\chi}^\pm}> 70$ GeV,
and also solutions which do not satisfy the bounds from the branching 
ratio on $B \to X_s \gamma$. 
The chargino contribution to $\cse$ decreases for larger values of 
$m_{\tilde{t}_1}$ and we therefore keep it to its minimal value 
$m_{\tilde{t}_1}= 70$ GeV. 
We have chosen a stop mixing angle $\theta_{\tilde{t}}= \pm 2 \pi/5$, i.e. 
the light stop $\tilde{t}_1=\cos\theta_{\tilde{t}} \tilde{t}_L+
\sin\theta_{\tilde{t}} \tilde{t}_R$ is  almost right handed.

For small $\tan\beta$, for which we again take $\tan\beta=2$, we find that 
the ratio $R_7$ remains positive, i.e. $\cse <0$, and lies within the 
experimentally allowed 
bounds from $B \to X_s \gamma$, and the other two ratios are in the range   
$0.98< R_9<1.07$ and $0.79< R_{10}< 1.15$.
For large $\tan\beta$, taken to be 30, just as in the SUGRA models
discussed earlier, $\cse$ changes 
sign ($R_7 <0$). The ratios $R_9$ and $R_{10}$ are again always positive
but now $R_9$ is almost identical to 1, and $R_{10}$ tends to lie below 
the SM-value. Numerically, we find the ranges
$0.99< R_9<1$, $0.93< R_{10}< 1.02$. 
The maximal (minimal) value of $\tan\beta$ found for $R_7 >0(<0)$
is 5 (20). In contrast, a no mixing choice 
$\theta_{\tilde{t}} =\pm \pi/2$ or $\tilde{t}_1\simeq \tilde{t}_R$ 
yields, for both $\tan\beta=2$ and 30, $\cse <0$ or equivalently $R_7 >0$.

In general, in the MFV model, SUSY effects on $\cn$ and $\ct$ are 
much smaller than the corresponding one on
$\cse$. A large value of $\tan \beta$ helps
$\cse$ to satisfy the $B\to X_s \gamma$ bounds but admits a sign opposite
to the one in the SM.
Dominant SUSY  contributions to $\cn$ and $\ct$ are due to the charged Higgs 
exchange and are  
suppressed as $\sim 1/\tan^2\beta$, for large $\tan \beta$.
 Chargino effects in  $\cn$ and $\ct$ increase for larger values of the 
ratio $M_2/|\mu_{susy}| > 1$. Using the central values of 
the parameters and the LCSR form 
factors, the maximal non-resonant branching ratios in the MFV are found
for the ratios $R_7=-1.2$, $R_9=1.0$ and $R_{10}=1.02$: 
${\cal{B}}_{nr}^{max MFV}(B \to K \mu^{+} \mu^{-})=7.5 \times 10^{-7}$ and
${\cal{B}}_{nr}^{max MFV}(B \to K^* \mu^{+} \mu^{-})=3.2 \times 10^{-6}$.
While  larger than the corresponding branching ratios in the SM,
they are compatible with the present experimental bounds 
\cite{cdfexcl,cleoexcl}. Our findings in the MFV-SUSY model  are very 
similar to 
the SUGRA case and in agreement with \cite{hw97} for the inclusive decays. 
As the values of $R_i$ for the maximal non-resonant branching ratios in 
the MFV model are almost identical to their SUGRA-model counterparts
given in Eq.~(\ref{eq:sugrar}),
for which we have shown the dilepton invariant mass spectra and FB-asymmetry,
we refrain from showing the corresponding figures for the MFV case.
\subsection{$B \to (K,K^*) \ell^+ \ell^-$ in the MIA Approach}
The minimal insertion approach aims at including all 
possible squark mixing effects in a model independent way. Choosing a 
$q,\tilde{q}$ basis where the $q-\tilde{q}-\tilde{\chi}^0$ 
and $q-\tilde{q}-\tilde{g}$
couplings are flavor diagonal, flavor changes are incorporated by a
non-diagonal mass insertion in the  
$\tilde{q}$ propagator, which can be parametrized as ($A,B=$Left, Right)
\cite{hkr86}
\begin{eqnarray}
(\delta_{ij}^{up,down})_{A,B}=
\frac{(m_{ij}^{up,down})^2_{A,B}}{m_{\tilde{q}}^2}\; ,
\end{eqnarray}
where $(m_{ij}^{up,down})^2_{A,B}$ are the off-diagonal elements of the
up(down) squark mass squared matrices that mix flavor $i$ and $j$, for
both the right- and left-handed scalars, and $m_{\tilde{q}}^2$ is the
average squark mass squared. The sfermion propagators are 
expanded in terms of the $\delta$s. The Wilson coefficients have 
the following  structure ($k=7,9,10$):
\begin{equation}
{\cal{C}}_k={\cal{C}}_k^{SM}+{\cal{C}}_k^{diag}+{\cal{C}}_k^{MIA},
\label{eq:coeffMIA}
\end{equation}
 where
${\cal{C}}^{MIA}$ is given in terms of $(\delta_{ij}^{up,down})^2_{A,B}$ up 
to two mass insertions \cite{LMSS99}, and ${\cal{C}}_k^{diag}$ being the
SUSY contribution in the basis where only flavor-diagonal contributions
are allowed. It is tacitly assumed that the
$\delta$s are small and this defines the theoretical consistency of
this approach which has to be checked {\it a posteriori}. 

The MIA-SUSY approach has been recently used in the analysis of the decays
$B \to X_s \ell^+ \ell^-$ \cite{LMSS99}, taking into account the
present bounds on the coefficient $\cse(m_B)$ following from the
decay $B \to X_s \gamma$. The other two coefficients ${\cal{C}}_{9}^{MIA}$
and ${\cal{C}}_{10}^{MIA}$ are calculated by scanning over the allowed 
supersymmetric parameter space \cite{LMSS99}.
For $\mu_{susy} \simeq -160$ GeV, $m_{\tilde{g}} \simeq m_{\tilde{q}}
\simeq 250$ GeV, $m_{\tilde{t}_1}=90$ GeV, $m_{\tilde{\nu}} \simeq 50$ GeV, 
these coefficients  are expressed as:
 \begin{eqnarray}
{\cal{C}}_{9}^{MIA}(m_B) &=& -1.2(\delta^u_{23})_{LL} + 0.69
(\delta^u_{23})_{LR} -0.51(\delta^d_{23})_{LL}~, \nonumber\\
{\cal{C}}_{10}^{MIA} &=& 1.75(\delta^u_{23})_{LL} -8.25
(\delta^u_{23})_{LR}~.
 \label{eq:lmss}
\end{eqnarray}
Of these, the mass insertions $(\delta^d_{23})_{LL}$ and 
$(\delta^u_{23})_{LL}$ are related by a CKM rotation and the bound on one
implies a similar bound on the other. 
One may have marked enhancement or depletion in the
branching ratios for the decay $B \to X_s \ell^+ \ell^-$.
Note also the large numerical coefficient of $(\delta^u_{23})_{LR}$
in the expression for ${\cal{C}}_{10}^{MIA}$. For the parameters for which
Eq.~(\ref{eq:lmss}) holds, the diagonal-SUSY contributions to $\cn$ and 
$\ct$ are: $\cn^{diag}(m_B)=-0.35, ~\ct^{diag}=-0.27$.  
 Depending on the value of
$(\delta^u_{23})_{LR}$ and $(\delta^u_{23})_{LL}$, the coefficient  
${\cal{C}}_{10}^{MIA}$
may easily overcome the SM- and the diagonal-MSSM-contributions in this   
coefficient, changing the overall sign of the FB-asymmetry. This
feature  is a
marked difference between this scenario and the competing ones, namely SUGRA
and MFV, where $C_{10}$ remains close to the SM 
value (see Table 1). This feature has been noted 
already in \cite{LMSS99} in the
context of the FB-asymmetry in the inclusive decay $B \to X_s \ell^+ \ell^-$.

 To maximize the effects in this general flavor-violating supersymmetric 
context, several special cases have been studied in Ref.~\cite{LMSS99} in 
detail. We shall discuss the following three scenarios from this 
work:\footnote{The specific values given above for the mass insertion 
parameter $(\delta^d_{23})_{LL}$ have been kindly
provided to us by Ignazio Scimemi. We also draw attention to several
misprints in the Tables given in \cite{LMSS99} and trust that an Erratum
is being issued by the authors of Ref.~\cite{LMSS99}.}
 \begin{enumerate}
\item ''Best enhancement scenario" for the branching ratio ${\cal B}(B \to 
X_s \ell^+ \ell^-)$, which corresponds to the choice $\cse=0.445$, 
$(\delta^d_{23})_{LL} = (\delta^u_{23})_{LL}=-0.5$ and
$(\delta^u_{23})_{LR} =0.9$; 
\item ''Best enhancement scenario with $\cse <0$", corresponding to
using $\cse=-0.445$, $(\delta^d_{23})_{LL} =-0.5, 
~(\delta^u_{23})_{LL}=-0.1$ and $(\delta^u_{23})_{LR} =0.9$;
\item ''Best depression scenario", corresponding to $\cse=-0.25$,
$(\delta^d_{23})_{LL} =0.5,
~(\delta^u_{23})_{LL}=0.1$ and $(\delta^u_{23})_{LR} =-0.6$.
\end{enumerate}
With these choices, drastic
effects in the branching ratios and the
FB-asymmetry have been predicted for the decays $B \to X_s \ell^+ \ell^-$, 
as  
displayed in Figs.\ 5--8 in Ref.~\cite{LMSS99}. To wit, in the first 
scenario listed above, enhancements
as large as a factor $5$ are admissible in ${\cal B}(B\to X_s e^+ e^-)$ and
even higher, 6.5, in $B \to X_s \mu^+ \mu^-$.

 We shall largely follow this analysis here in discussing the decay
characteristics of the exclusive decays $B \to (K,K^*) \ell^+ \ell^-$
but would like to add a dissenting remark concerning the coefficient 
$\cse(m_B)$. We recall
that the extremal values used for $\cse(m_B)$ in     
\cite{LMSS99} correspond to using the 99\% C.L. limits on ${\cal B}(B    
\to X_s \gamma)$, which give the bounds $0.252 < \vert \cse  
\vert < 0.445$ in the NLO approximation. This procedure allows a
much larger range for the ratio $R_7$ than the one given in 
Eq.~(\ref{eq:R7bounds}), which is then partly reflected 
in the branching ratios for $B \to X_s \ell^+ \ell^-$.

We argue that even with this more restricted range of $\cse$,
the two ''Best enhancement scenarios for $B \to X_s \ell^+ \ell^-$" of
Ref.~\cite{LMSS99} alluded to above give too large branching ratios for the 
exclusive
decays being studied here. To be specific, in the first scenario, 
the parameters given above translate into 
$R_9=1.26$ and $R_{10}=2.84$.\footnote{
 We neglect the effect from the RG running from $\mu=m_B$ (used in
\cite{LMSS99}) to $\mu=m_{b, pole}$ used by us.}
The central values of the form factors calculated 
here in the LCSR approach then lead to the following branching ratio:
${\cal{B}}_{nr}^{max, MIA}(B \to K^* \mu^{+} \mu^{-})=11.5 \times 10^{-6}$,
which is approximately 3 times larger than the recent CDF (90\% C.L.) upper 
limit on this quantity \cite{cdfexcl},
\begin{equation}
{\cal B}(B^0 \to K^{*0} \mu^{+} \mu^{-})< 4.0 \times 10^{-6}~.
\label{eq:cdfbound}
\end{equation} 
The $B \to K$ transition in this scenario is likewise enhanced, yielding a 
branching ratio  
${\cal{B}}_{nr}^{max, MIA}(B \to K \mu^{+} \mu^{-})=3.2 \times 10^{-6}$,
which is typically a factor 5 larger than the  
SM branching ratio, but still compatible with the
experimental upper limit, ${\cal{B}}(B^+ \to K^+ \mu^{+} 
\mu^{-}) <5.2 \cdot 10^{-6}$ \cite{cdfexcl}. Hence, the 
present experimental upper bound on $B \to K^* \mu^+ \mu^-$ provides
non-trivial bounds on $C_9$ and $C_{10}$, equivalently on $R_9$ and 
$R_{10}$, which we now proceed to work out.

\subsection{Bounds on $\cn$ and $\ct$ from present data}
The branching ratios $B \to (K,K^*) \ell^+ \ell^-$ can be expressed as  
quadratic equations in the coefficients $\cse$, $\cn$ and $\ct$.
Given the branching ratios (equivalently upper bounds),  
these equations can be solved numerically and yield
the allowed contours in the $\cn$-$\ct$ plane. For working out the 
constraints, we use the experimental bound in 
Eq.~(\ref{eq:cdfbound}) and the following expression which follows from
Eq.~(\ref{eq:dwbvll}):
\begin{equation}
{\cal{B}}(B \to K^* \mu^+ \mu^-) 
=a^{(nr)}_{K^*} |\cse|^2 +b^{(nr)}_{K^*} |\cn|^2+c^{(nr)}_{K^*} |\ct|^2 
+d^{(nr)}_{K^*} \cse \cn +e^{(nr)}_{K^*} \cse +f^{(nr)}_{K^*} \cn 
+g^{(nr)}_{K^*}~. \label{eq:quadratic}
\end{equation}
The coefficients $a^{(nr)}_{K^*},...,g^{(nr)}_{K^*}$ are tabulated in 
Table~\ref{tab:bkstar}, using the central values of the $B \to K^*$ 
form factors in Table~\ref{tab:p1} and the
maximum and minimum values of the same given in Tables~\ref{tab:p2}
and \ref{tab:p3}, respectively. Of these, the coefficients $b^{(nr)}_{K^*}$
and $c^{(nr)}_{K^*}$ coincide if one neglects the $\ell^\pm$-masses.
The superscript on these coefficients
is a reminder that only non-resonant contributions are included.
\begin{table}[t]
        \begin{center}
        \begin{tabular}{|l|c|c|c|c|c|c|c|}
\hline
    \multicolumn{1}{|l|}{\mbox{}}
      & \multicolumn{1}{|c|}{$a^{(nr)}_{K^*}$}
      & \multicolumn{1}{|c|}{$b^{(nr)}_{K^*}$} 
      & \multicolumn{1}{|c|}{$c^{(nr)}_{K^*}$}
      & \multicolumn{1}{|c|}{$d^{(nr)}_{K^*}$} 
      & \multicolumn{1}{|c|}{$e^{(nr)}_{K^*}$}
      & \multicolumn{1}{|c|}{$f^{(nr)}_{K^*}$} 
      & \multicolumn{1}{|c|}{$g^{(nr)}_{K^*}$} \\
        \hline
     FF(central) &21.295  &0.502&0.500&3.530&1.434&0.413&0.148\\
     FF(max)   &28.183  &0.630&0.633&4.577&1.859&0.520&0.183\\
     FF(min)   &16.795  &0.417&0.416&2.864&1.164&0.343&0.125\\
\hline
\end{tabular}
\end{center}
\caption{\it Coefficients of the non resonant branching ratio 
${\cal{B}}( B \to K^* \mu^+ \mu^-)$ in units of $10^{-7}$ in the
decomposition as in Eq.~(\ref{eq:quadratic}), integrated over the 
full $q^2$ range for different sets of form factors given in 
Tabs.~\ref{tab:p1}-\ref{tab:p3}.}
\label{tab:bkstar}
\end{table}

\begin{figure}[p]
\vskip 0.0truein
\centerline{\epsfysize=3.3in
{\epsffile{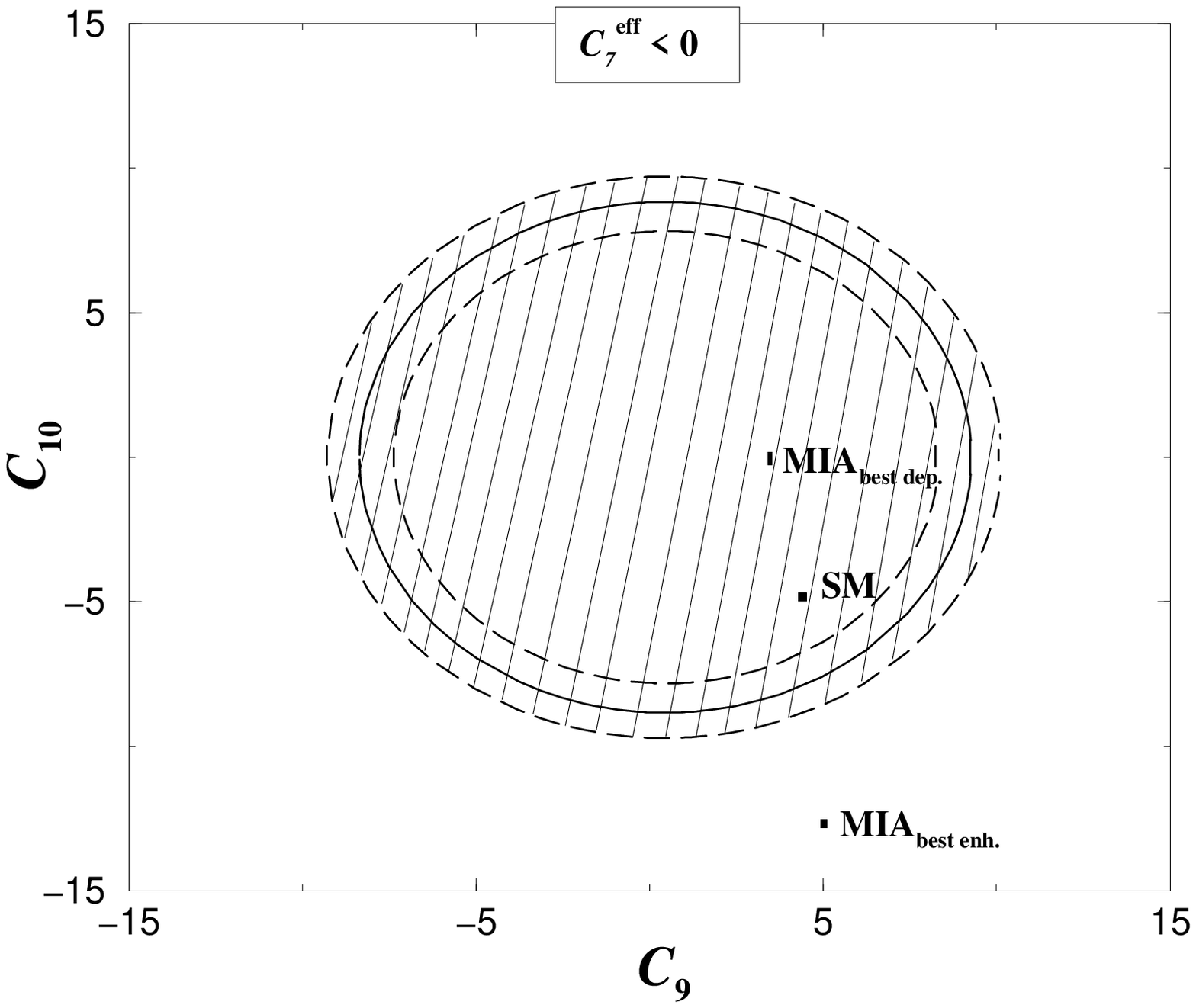}}}
\vskip 0.0truein
\caption[]{ \it Bounds on the coefficients $C_9(m_B)$ and $C_{10}$ resulting
from the experimental upper bound ${\cal B}(B^0 \to K^{*0} \mu^{+}   
\mu^{-})< 4.0 \times 10^{-6}$ (at 90\%
C.L.) \protect\cite{cdfexcl} and $\cse(\mu=4.8 ~\mbox{GeV})=-0.249$ from
the bounds given in Eq.~(\protect\ref{eq:c7lla}).The SM-point   
and two representative points in the SUSY-MIA approach from
Ref.~\protect\cite{LMSS99} are also shown. The three curves correspond to
using the central values of the form factors (solid curve), the
minimum (outer dashed curve) and maximum (inner dashed curve) allowed
values discussed in Sec.~3.}
\label{fig:cdfsm}
$$\epsfysize=3.3in
{\epsffile{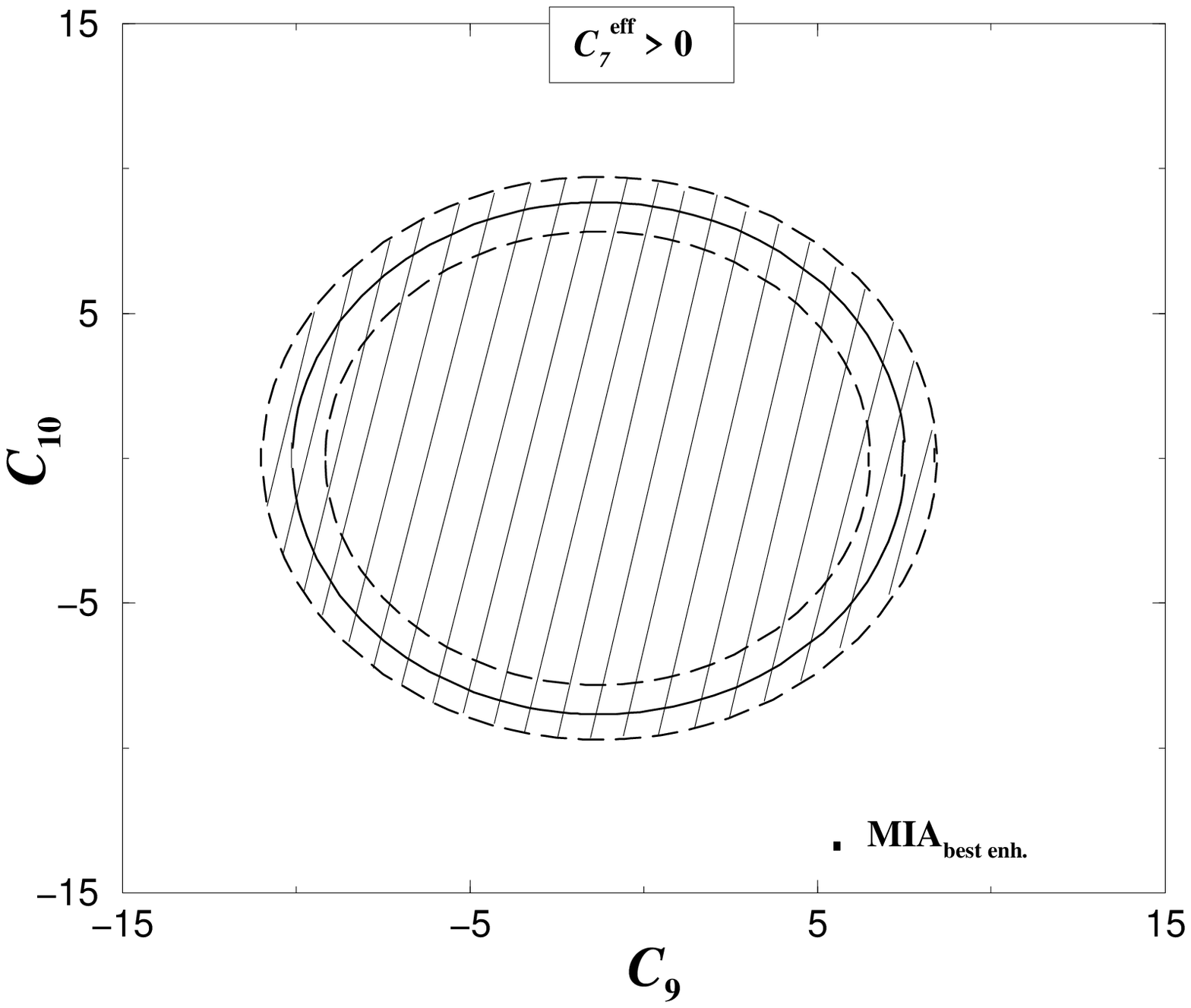}}
$$
\vspace*{-28pt}
\caption[]{ \it The same as Fig.~\protect\ref{fig:cdfsm} but for the
solution with $\cse=0.249$. The point $MIA_{best}$ corresponds to the
"best enhancement scenario" of Ref.~\protect\cite{LMSS99}, discussed
in the text.}
\label{fig:cdfnp}
\end{figure}

\begin{figure}[p]
\vskip 0.0truein
\centerline{\epsfysize=3.3in
{\epsffile{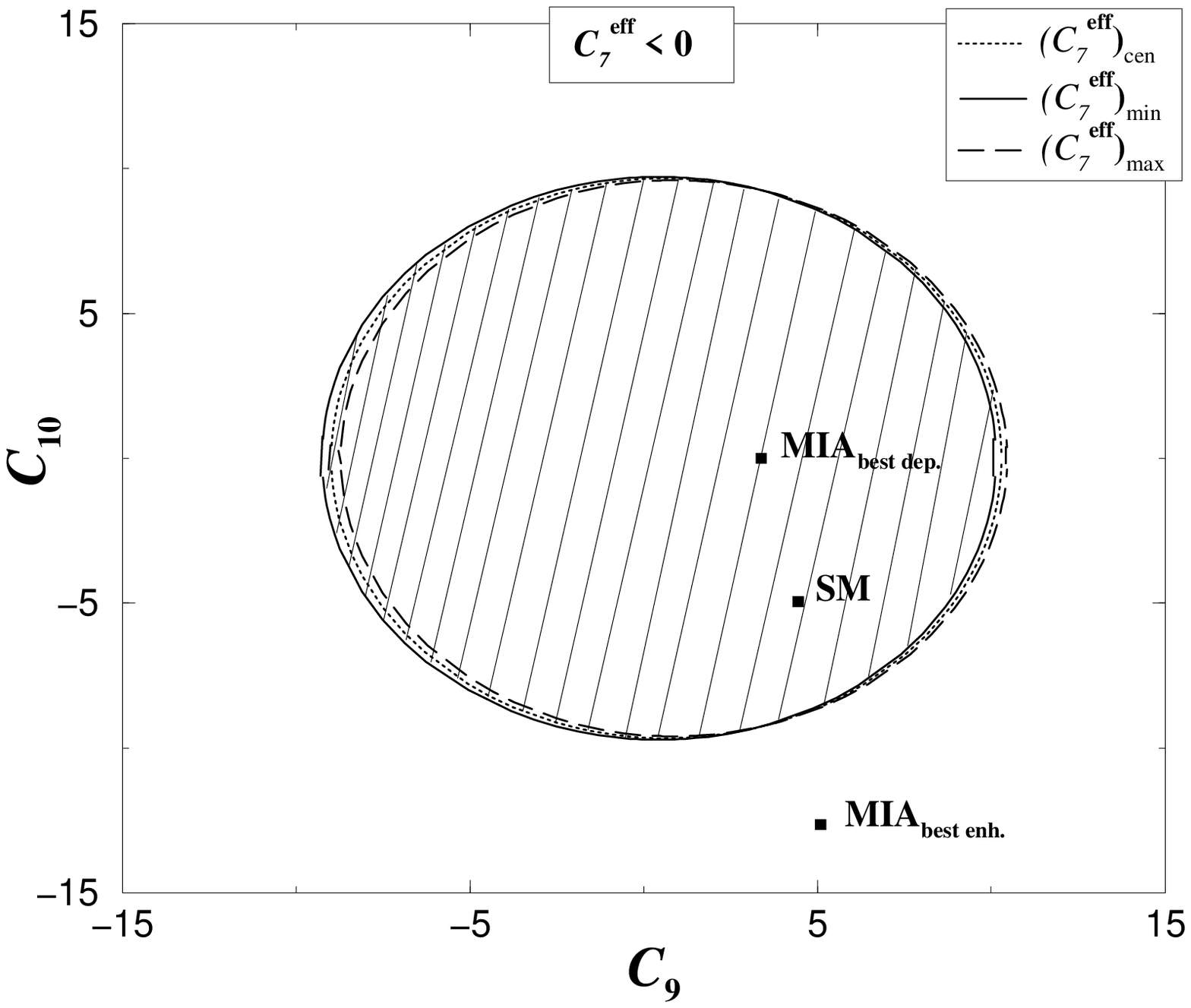}}}
\vskip 0.0truein
\caption[]{ \it The same as Fig.~\protect\ref{fig:cdfsm}, but showing
  the dependence of the bounds on the experimentally allowed range for
  $\vert\cse\vert$, 
$0.249 \leq \vert\cse\vert \leq 0.374$, with the form factors 
fixed to their minimum values given in Table~\protect\ref{tab:p3}.}
\label{fig:cdfsmC7}
$$\epsfysize=3.3in
{\epsffile{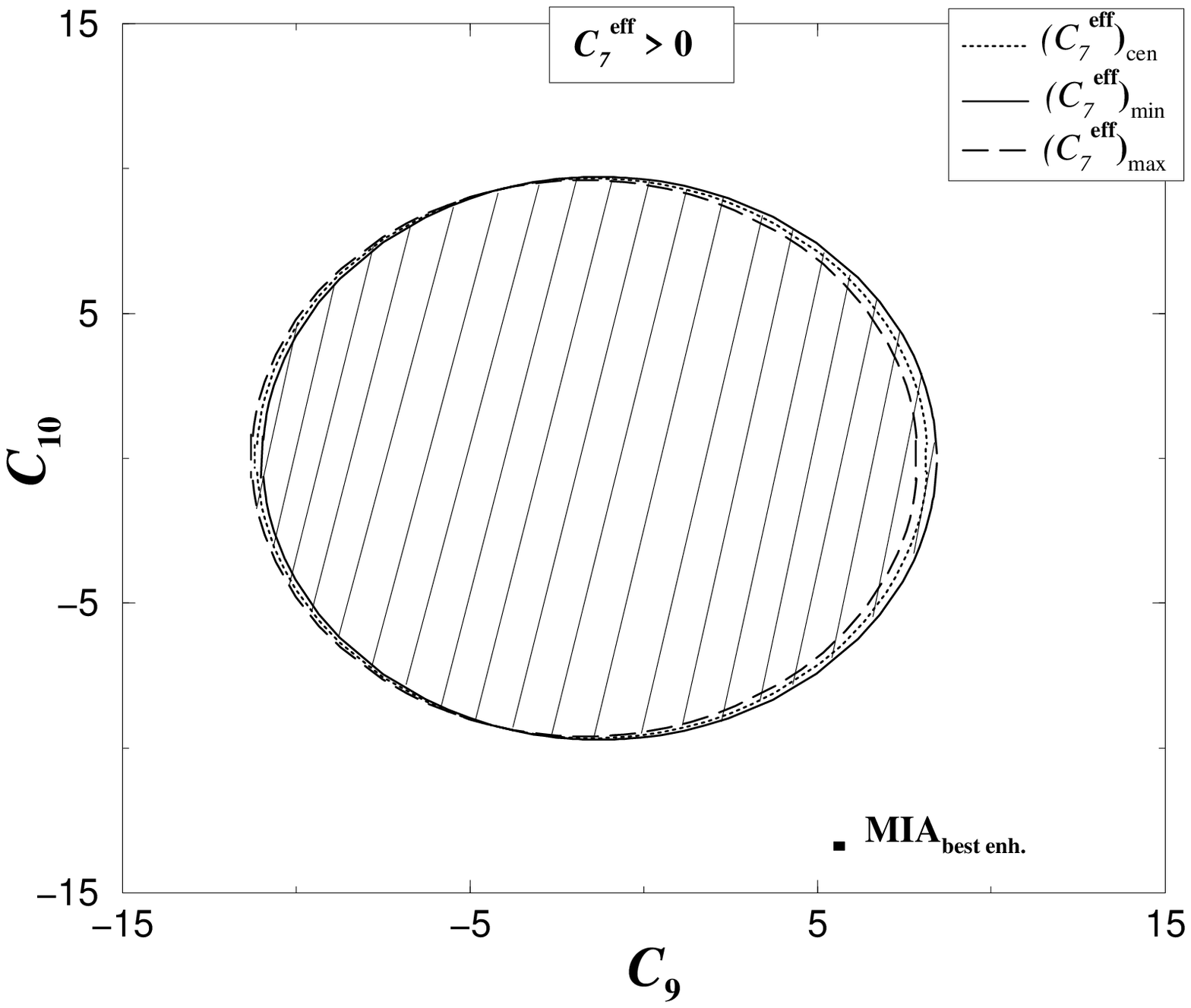}}
$$
\vspace*{-28pt}
\caption[]{ \it The same as Fig.~\protect\ref{fig:cdfsmC7}, but
with $\cse >0$.}
\label{fig:cdfnpC7}
\end{figure}

The quadratic equation in (\ref{eq:quadratic}) is solved numerically for 
the two distinct situations $\cse <0$ (SM-like) and $\cse >0$ (new physics 
scenario) in the experimentally allowed range for $\cse$ given in 
Eq.~(\ref{eq:c7lla}). The resulting 90\% C.L.
allowed contours are shown in Fig.~\ref{fig:cdfsm} and Fig.~\ref{fig:cdfnp},
respectively. The solid curves in these figures are obtained by using the 
central values of
the form factors and the inner and outer dashed curves represent the
maximal and minimal allowed values of the same, respectively.
Note that the loosest bounds emerge 
from the minimal allowed values of the form factors. Also, in working out 
the constraints shown in
these figures, we have fixed $\vert\cse\vert =\vert\cse_{min}\vert=0.249$
in the allowed range given in Eq.~(\ref{eq:c7lla}),
as this gives for both the cases ($\cse <0$ and $\cse >0$) the loosest
bounds on $\cn$ and $\ct$. This can be seen in
Figs.~\ref{fig:cdfsmC7} and Fig.~\ref{fig:cdfnpC7} drawn for $\cse<0$ and
$\cse>0$, respectively, where we 
show the dependence of the bounds in the $\cn$-$\ct$ plane on the 
experimentally allowed range for $\vert\cse\vert$ given in 
Eq.~(\ref{eq:c7lla}). In these figures, we use the minimum values of 
the form factors given in Table~\ref{tab:p3} for reasons given above. 
 In Figs.~\ref{fig:cdfsm} and \ref{fig:cdfsmC7}, 
we also show the SM-point (see Table 1) and the
SUSY-MIA points for the ''Best enhancement scenario with $\cse 
<0$", corresponding to $\cn(m_B)=5.0, ~\ct=-12.5$, and the ''Best 
depression scenario with $\cse <0$", corresponding to $\cn(m_B)=3.2, 
~\ct=0.2$, \cite{LMSS99}. We note that the ''Best enhancement scenario
with $\cse <0$" is ruled out by data.  
 The other MIA-SUSY point, as well as the SM, are both well within the
experimental bound.  
The SUSY-MIA point corresponding to the ''Best enhancement 
scenario with $\cse >0$" of Ref.~\cite{LMSS99} is shown in the $\cn$--$\ct$ 
plane in  Figs.~\ref{fig:cdfnp} and \ref{fig:cdfnpC7}. 
This corresponds to the point $\cn(m_B)=5.5, ~\ct=-13.2$.
 As anticipated, this ''Best 
enhancement scenario with $\cse >0$" is convincingly ruled out by the
experimental upper bound on ${\cal B}(B \to K^* \mu^+\mu^-)$.
The analysis shown in Figs.~\ref{fig:cdfsm}-\ref{fig:cdfnpC7} holds for 
all models 
discussed here in this paper in which the SD-physics can be encoded in 
terms of the three real Wilson coefficients $\cse$, $\cn$ and $\ct$.
The point we wish to stress is that existing data on
$B \to K^* \mu^+ \mu^-$, in conjunction with the branching ratio
${\cal B}(B \to X_s \gamma)$, provides  non-trivial constraints on $\cn$ 
and $\ct$. 

Illustrative examples of the dilepton invariant mass spectrum in the
decays $B \to K \mu^+ \mu^-$ and $B \to K^* \mu^+ \mu^-$ in the MIA 
approach are shown in Figs.~\ref{fig:BKsusy} and \ref{fig:BKstsusy},
respectively. They have been calculated for the following values:
\begin{equation}
R_7=\pm 0.83, ~~R_{9}=0.92, ~~~R_{10}=1.61 ~,
\label{eq:MIAplots}
\end{equation}
which are allowed by the present experimental bounds. The 
characteristic difference in this case, as compared to the SUGRA and 
MFV-SUSY models, lies in the significantly enhanced value of $\ct$.   

As already mentioned, a characteristic of the MIA approach is that 
 the sign of $\ct$ ($C_{10}^{SM}<0$) depends on the quantities
$(\delta^u_{23})_{LR}$ and $(\delta^u_{23})_{LL}$. In particular, the
large number in front of $(\delta^u_{23})_{LR}$ in $C_{10}$,
 obtained for the specific values of the SUSY parameter space, could  
change the sign of this Wilson coefficient. This has 
 no effect on the dilepton invariant mass distributions,
as they depend quadratically on $C_{10}$, 
but it would change the sign of $A_{FB}$ in $B \to K^* \ell^+ \ell^-$. To 
illustrate 
this, we use the parameters close to the so-called "Best depression" 
scenario \cite{LMSS99}, corresponding to the following values
\begin{equation} 
R_7=\pm 0.83, ~~R_9=0.79, ~~R_{10}=-0.38~,
\label{eq:bestdepr}
\end{equation}
 and plot the resulting normalized FB asymmetry in 
Fig.~\ref{fig:BKstAFBsusy}. The positive FB-asymmetry in $B \to K^* \ell^+
\ell^-$ (as well as in $B \to X_s \ell^+ \ell^-$ shown in 
\cite{LMSS99}) for the dilepton invariant mass below the resonant $J/\psi$ 
region is rather unique, as none of
the other models considered here (SM, SUGRA and MFV) admit solutions
with positive $C_{10}$. 

Finally, to facilitate a model independent determination of the coefficients
$\cse$, $C_9$, and $C_{10}$ from the decays $B \to (K,K^*) \ell^+ 
\ell^-$, we write down a   
parametrization of the partially integrated branching ratios and FBA in the 
low $s$ region. Using, for the sake of definiteness,
$s_{min}=0.25~\mbox{GeV}^2$, $s_{max}=8.0~\mbox{GeV}^2$,
the partial branching ratios $\Delta {\cal{B}}_X$ and the corresponding
FB-asymmetry $\Delta {\cal{A}}_{FB}$ can be expressed as 
($X=K,K^*$):
 \begin{eqnarray}
\Delta {\cal{B}}_X &\equiv&
\int_{s_{min}}^{s_{max}} \d s \frac{\d {\cal{B}}(B \to X \mu^+ \mu^-)}{\d s} \\
&=&a_X |\cse|^2 +b_X |\cn|^2+c_X |\ct|^2 +d_X \cse \cn +e_X \cse +f_X \cn +g_X \label{eq:coeffbr}\\
\Delta {\cal{A}}_{FB}&\equiv& \tau_B
\int_{s_{min}}^{s_{max}} \d s 
\frac{\d {\cal{A}}_{FB}(B \to K^* \mu^+ \mu^-)}{\d s}
=\ct (h_X \cse +j_X \cn +k_X)
\label{eq:coefffba}
\end{eqnarray}
\begin{table}[t]
        \begin{center}
        \begin{tabular}{|l|c|c|c|c|c|c|c|c|c|c|}
\hline
    \multicolumn{1}{|l|}{\mbox{}}
      & \multicolumn{1}{|c|}{$a$}
      & \multicolumn{1}{|c|}{$b$} 
      & \multicolumn{1}{|c|}{$c$}
      & \multicolumn{1}{|c|}{$d$} 
      & \multicolumn{1}{|c|}{$e$}
      & \multicolumn{1}{|c|}{$f$} 
      & \multicolumn{1}{|c|}{$g$} 
      & \multicolumn{1}{|c|}{$h$}
      & \multicolumn{1}{|c|}{$j$} 
      & \multicolumn{1}{|c|}{$k$} \\
        \hline
     $K$   & 0.193  &0.068&0.068&0.230&0.163&0.097& 0.045&-&-&-  \\
     $K^*$ & 13.119&0.197&0.196&1.760&0.995&0.236&0.083&0.943&0.089&0.061  \\
\hline
\end{tabular}
\end{center}
\caption{\it Coefficients in units of $10^{-7}$ defined in 
Eqs.~(\ref{eq:coeffbr}) and (\ref{eq:coefffba}) in the KS prescription 
\cite{ks96}.} \label{tab:coeff}
\end{table}
Numerical values of the coefficients are given in Table \ref{tab:coeff}.
They have been obtained by using the central values of the form factors
and other parameters given in Table \ref{tab:p1} and Table \ref{parameters},
respectively. Specifying a model by the effective coefficients
$\cse(m_B)$, $\cne(m_B)$ and $C_{10}$ enables one to
obtain readily the predictions for $\Delta {\cal{B}}_X$ and $\Delta 
{\cal{A}}$ in this model. In the SM, we estimate
$\Delta {\cal{B}}_K=2.90 \cdot 10^{-7}$,
$\Delta {\cal{B}}_{K^*}=7.67 \cdot 10^{-7}$ and
$\Delta {\cal{A}}_{FB}=-0.71 \cdot 10^{-7}$,
yielding $\Delta \bar{{\cal{A}}}_{FB}
=\Delta {\cal{A}}_{FB}/\Delta {\cal{B}}_{K^*}=-9.2 \%$. The branching ratios
for the decays $B \to (K,K^*) e^+ e^-$ are practically identical.
Typical theoretical errors on these quantities, obtained by varying the 
form factors and the parameters $m_t$, $m_b$, $\mu$ and $\Lambda_{QCD}$
in the ranges discussed earlier and adding the individual errors in 
quadrature are $\pm 
30\%$ for $\Delta {\cal{B}}$ and  $\pm 38\%$ for $\Delta {\cal{A}}$.
However, the branching ratios and the FB-asymmetry may be significantly 
enhanced  (or depressed) in some variants of the supersymmetric models 
discussed.
 With $O(10^8)$   
$B\bar{B}$ events anticipated at the B-factories and HERA-B, and much
higher yields at the Tevatron and
LHC experiments, these rates and asymmetries will allow precision tests of
the SM and may indicate the presence of  new physics.

\section{Summary and Concluding Remarks}
 Before summarizing our results, we would like to comment
on the contributions from the helicity-flipped counter-parts of 
the SM operators ${\cal{O}}_{7}, ~{\cal{O}}_{9}$ and ${\cal{O}}_{10}$:
\begin{eqnarray}
{\cal{O}}^{\prime}_7   &=&\frac{e}{16 \pi^2} \bar{s} \sigma_{\mu \nu}
m_b L b F^{\mu \nu}~, \\
{\cal{O}}^{\prime}_9   &=& \frac{e^2}{16 \pi^2} 
\bar{s}_R \gamma^\mu b_R \bar{\ell} \gamma_\mu \ell~, \\
{\cal{O}}^{\prime}_{10}&=& \frac{e^2}{16 \pi^2} 
\bar{s}_R \gamma^\mu b_R \bar{\ell} \gamma_\mu \gamma_5 \ell ~.
\end{eqnarray}
In an enlarged operator basis including these and the SM-operators, the 
various distributions 
for the decays of interest can be obtained from the substitutions 
$\c_i \to \c_i+ \c_i^{\prime}$ ($i=7,9,10$) in the matrix elements and the 
auxiliary 
functions Eqs.~(\ref{eq:aux1})--(\ref{eq:aux2}) for $B \to K$, and for 
$B \to K^*$ in the terms which are proportional to the form factors 
$V$ and $T_1$. In the remainder of the $B \to K^*$ amplitude,
the contribution of the helicity-flipped operators
enters with the opposite sign, i.e., $\c_i \to \c_i- \c_i^{\prime}$.

We note that in all models with minimal flavor violation, like the
SM, 2HDM,  and MFV, the contributions of the flipped operators 
$\o^\prime_{7,9,10}$
vanish in the $m_s \to 0$ limit.
In the general non-diagonal MSSM scenarios, there are finite contributions
even for a vanishing $s$-quark mass due to the neutralino-gluino-down-squark 
loops.
However, under the assumption that no large cancellations happen, we can 
conclude from the data on  
${\cal{B}}(B \to X_s \gamma)$ which bounds $  |\cse + \c_7^{{\rm 
eff}\prime}|^2$ 
that $\c_7^{{\rm eff} \prime}$ must be small compared to $\cse$. Further, 
neglecting box diagrams, the helicity structure of the (penguin)-loops 
responsible for 
$\c_{9,10}^{\prime}$ can be related to the ones of the flipped photon 
penguin $\c_7^{{\rm eff}\prime}$ and hence is suppressed as well.
We also note that we have neglected the effects
of the neutral Higgs exchanges, which may lead to some inaccuracies for the
decay $B \to (K,K^*) \tau^+ \tau^-$ in some parts of the SUSY parameter
space. They are insignificant for the decays involving the $(K,K^*) 
\mu^+\mu^-$ and $(K,K^*)e^+e^-$ states, where most of the experimental
searches will be concentrated. 

We summarize our results: We have undertaken an improved calculation of
the form factors in the decays $B \to (K,K^*) \ell^+ \ell^-$ in the light
cone QCD sum rule approach. Using this framework,
we have calculated the partial branching ratios, dilepton invariant mass
spectra and the forward-backward asymmetry for these decays in the context of
the SM. We have also undertaken a comparative study of the phenomenological
profiles of these decays in a number of supersymmetric models. These
include the SUGRA models, minimal-flavor-violation SUSY model, and a general
flavor-violating SUSY framework using the mass insertion approximation.
The role of the forward-backward 
asymmetry in the decays $B \to K^* \ell^+ \ell^-$ in searching for new 
physics is emphasized. We show that the large-$(\tan\beta)$ solution in
the SUGRA models, but also some parameter space of the MIA model, yield
FB-asymmetries, which are strikingly different from the SM. 
 In particular, the value of the dilepton invariant 
mass for which the FB-asymmetry may become zero, $s_0$, may 
provide a precision test of the SM. A simple analytic expression for  
$s_0$ is derived, and we have argued that the form factor 
dependence in $s_0$ cancels in  the large energy expansion approximation.  
We have analyzed the present data on $B \to X_s \gamma$ and existing 
limits on the decays $B \to (K,K^*) \ell^+ \ell^-$ to put bounds on the
coefficients $\cn$ and $C_{10}$. While these bounds do not yet 
probe the SM,
they do provide non-trivial constraints on extensions of the SM.
In particular, the ''Best enhancement SUSY-MIA scenarios "for 
the branching 
ratios ${\cal B}(B \to X_s \ell^+ \ell^-)$, shown for some chosen 
supersymmetric parameters  in 
Ref.~\cite{LMSS99}, are ruled out by the existing upper limit on
the exclusive branching ratio ${\cal B}(B^0 \to K^{*0} \mu^+ \mu^-)$
\cite{cdfexcl}. Finally, we show the dilepton mass spectra and the 
FB-asymmetry for illustrative values of the supersymmetric parameters and
argue that the decays $B \to (K,K^*) \ell^+ \ell^-$ hold great promise
in unraveling new physics.

\section*{Acknowledgements}

A.A.\ would like to acknowledge helpful communication with Ignazio
Scimemi on the work reported in \cite{LMSS99}.
G.H.\ would like to thank Frank Kr\"uger and Tilman Plehn for useful  
discussions, and the Fermilab theory group for the hospitality during
her stay, where a part of this work has been done.
P.B.\ is supported  by the Deutsche Forschungsgemeinschaft (DFG) through a 
Heisenberg fellowship.
L.T.H.\ would like to thank the Alexander von Humboldt Stiftung 
for financial support.  This project is 
partially supported by the EEC-TMR Program, Contract N.~FMRX-CT98-0169.


\end{document}